\documentclass[aps,pra,twocolumn,eqsecnum,amsmath,superscriptaddress,showpacs]{revtex4-1}
\usepackage{graphicx}
\usepackage{longtable}
\usepackage{epsfig}
\usepackage{dcolumn}
\usepackage{bm}
\usepackage{amssymb}
\usepackage{multirow}
\usepackage{times,color}
\usepackage{hyperref}
\usepackage{ulem}
\begin{document}

\title{Exact treatment of the magnetism-driven ferroelectricity
       in the one-dimensional compass model}

\author {Wen-Long You }
\affiliation{College of Physics, Optoelectronics and Energy, Soochow
University, Suzhou, Jiangsu 215006, People's Republic of China}

\author {Guang-Hua Liu}
\affiliation{Department of Physics, Tianjin Polytechnic University,
Tianjin 300387, People's Republic of China}

\author {    Peter Horsch }
\affiliation{Max-Planck-Institut f\"ur Festk\"orperforschung,
             Heisenbergstrasse 1, D-70569 Stuttgart, Germany }

\author {    Andrzej M. Ole\'s }
\affiliation{Max-Planck-Institut f\"ur Festk\"orperforschung,
             Heisenbergstrasse 1, D-70569 Stuttgart, Germany }
\affiliation{Marian Smoluchowski Institute of Physics, Jagellonian
             University, Reymonta 4, PL-30059 Krak\'ow, Poland }

\date{\today}

\begin{abstract}
We consider a class of one-dimensional compass models with antisymmetric
Dzyaloshinskii-Moriya exchange interaction in an external magnetic
field. Based on the exact solution derived by means of Jordan-Wigner
transformation, we study the excitation gap, spin correlations, ground
state degeneracy and critical properties at phase transitions. The phase
diagram at finite electric and magnetic field consists of three phases:
a ferromagnetic, a canted antiferromagnetic and a chiral phase.
Dzyaloshinskii-Moriya interaction induces an electrical polarization in
the ground state of the chiral phase, where the nonlocal string order
and special features of entanglement spectra arise, while strong chiral
correlations emerge at finite temperature in the other phases and are
controlled by a gap between the nonchiral ground state and the chiral
excitations. We further show that the magnetoelectric effects in all
phases disappear above a typical temperature corresponding to the total
bandwidth of the effective fermionic model. To this end we explore the
entropy, specific heat, magnetization, electric polarization, and the
magnetoelectric tensor at finite temperature.
We identify rather peculiar specific heat and polarization behavior of
the compass model which follows from highly frustrated interactions.
\end{abstract}


\maketitle

\section{Introduction}

In classical
electromagnetism, electric ($\vec{E}$) and magnetic ($\vec{H}$)
fields induce electric polarization ($\vec{P}$) and magnetization
($\vec{M}$) in matter, respectively, yet using $\vec{H}$ ($\vec{E}$)
to induce $\vec{P}$ ($\vec{M}$), the so-called magnetoelectric effect
(MEE), is a highly nontrivial issue \cite{Fiebig05}.
Recently, technological and theoretical progress triggered a
renaissance of the MEE, especially in multiferroic materials
\cite{Kimura03,Cheong07,Kitagawa10,Lee13}.
It was expected that the efficient control of magnetism in terms of the
electric field would have many potential applications in spintronics
and data storage technology.

There are several envisaged explanations for the magnetically induced
ferroelectricity. Among them, the exchange-striction mechanism
\cite{Choi08,Tokunaga08} and the inverse Dzyaloshinskii-Moriya (DM)
mechanism \cite{Katsura05} are main streams in accounting for
ferroelectricity in inversion symmetry breaking lattices. In the former
mechanism, the bonds become different between the parallel and the
antiparallel spins through exchange striction associated with symmetric
superexchange $(\vec{\sigma}_i\cdot\vec{\sigma}_j)$. As a result,
the electric polarization $\vec{P}$ is induced by the crystallographic
deformations in the direction of the chain. Such a mechanism was
originally introduced to explain the MEE in material Cr$_2$O$_3$
\cite{Date61} and was recently applied to other versatile systems
\cite{Kimura07,Pomjakushin09,Soda11}.

Besides, for the noncollinear spiral- or helimagnetic order, resulting
from antisymmetric magnetic frustration, the term
$(\vec{\sigma}_i\times\vec{\sigma}_j)$ is also a common way to activate
the inversion symmetry breaking. The electric polarization $\vec{P}$ is
thereby generated by the displacement of oppositely
charged ions as described by Tokura and Seki \cite{Tokura10},
\begin{eqnarray}
\vec{P}_{i} = \gamma \hat{e}_{ij} \times (\vec{\sigma}_i
\times \vec{\sigma}_j),
\label{polarization}
\end{eqnarray}
where $\hat{e}_{ij}$ is the unit vector connecting the neighboring
spins $\vec{\sigma}_i$ and $\vec{\sigma}_j$. The coupling coefficient
$\gamma$ of the cycloidal component is material-dependent
\cite{Sergienko06}, and its sign depends on the vector spin chirality.
This microscopic origin towards magnetism-induced ferroelectricity
constitutes the well-known inverse DM mechanism.
It was proposed that DM interaction induces the helimagnetic spin
ground state and ferroelectricity in Cu$_2$OSeO$_3$
\cite{Seki12,Adams12,Yang12}, cycloidal magnetic structure in
multiferroic BiFeO$_3$ \cite{Matsuda}. Note, incidentally, that there
is a plethora of experimentally accessible compounds where electron
spin resonance can be applied as a consequence of the presence of DM
interaction \cite{Yu11,Karimi11}.
The DM-like spin-orbit interaction in a single-crystal yttrium iron garnet was experimentally measured recently \cite{Zhang14}.
Meanwhile, a number of theoretical
papers was devoted to the effects of DM interaction in magnets
\cite{You11,Perk75,Zhong13,Brockmann13}.
Of interest are also studies that explore spin superfluidity as
a consequence of the non-collinear long-range order and the analogy
to Josepson effect in superconductors \cite{Chen14,Chen14a}.

The concept of magnetism-driven ferroelectricity naturally brings
significant attention to quantum spin systems. Among them, several
spin-1/2 chain materials appear to be the most straightforward but
conceptually important models, and have been extensively studied, as
for example Ca$_3$CoMnO$_6$ \cite{Choi08}, LiCu$_2$O$_2$ \cite{Park07},
LiCuVO$_4$ \cite{Schrettle08,Matiks09}, and CuCl$_2$ \cite{Seki10}.
In a realistic quantum wire, the superexchange interaction between
spins of transition metal ions depends on the details of crystal
structure, such as the bond length between magnetic ions and the angle
between the bonds connecting magnetic and ligand ions. The effect of
such variants must be addressed.

So far most literature focuses on
Heisenberg exchange interaction --- such models require approximations
and are not easy to solve completely. In order to obtain an unbiased
solution we address in this paper the problem of ferroelectricity in
the anisotropic exchange model which is exactly solvable. Our model is
a generalization of the one-dimensional (1D) compass model where vector
chiral correlations are introduced by the additional DM~interaction.
It features rather rich phase diagram that results from the interplay
of DM terms, external fields and frustrated compass-type exchange
interactions. We analyze the different symmetry breaking due to
magnetic and electric fields. Further we demonstrate in the fermionic
representation of the spin model that the gapless topological phase is
characterized by four distinct Fermi points, in contrast to the
conventional gapped phases or symmetry-protected phases with twofold
degenerate Fermi points. We identify the nonlocal string order in the
chiral phase accompanied by the finite electrical polarization.
In particular we demonstrate the change of the entanglement spectrum
between the canted N\'{e}el phase and the chiral phase.

The 1D compass model is a realization of the directional competing
interactions known from the two-dimensional (2D) compass model
\cite{Nus05,Dou05,Dor05,Brz10} on a chain with alternating Ising-like
interactions between $x$ and $z$ spin components on neighboring bonds.
Similar to the Kitaev model \cite{Kit06,Bas07}, the 2D and 1D compass
model are characterized by intrinsic frustration of interactions. An
exact solution of the 1D compass model shows that the ground state has
high degeneracy \cite{Brz07}, and a quantum phase transition (QPT)
occurs when anisotropic interactions pass through the isotropic point
\cite{You14}.

We have organized the paper into six sections. First, we introduce
the Hamiltonian of the 1D generalized compass model (GCM) with DM
interaction in Sec. \ref{sec:fermions}, and then present the procedure
to solve it exactly by employing Jordan-Wigner transformation in
absence of magnetic field, see Sec. \ref{sec:quasi}.
This solution is next used to evaluate various
correlation functions at finite electric field in Sec. \ref{sec:corr}.
The model in the magnetic field is analyzed in Sec. \ref{sec:HMF} and
the complete phase diagram is obtained when the electric and magnetic
field are varied. Several thermodynamic functions, such as the entropy
and the specific heat, and the MEEs are presented and
discussed in Sec. \ref{sec:Thermodynamic} and  \ref{sec:Magnetoelectric}.
Section \ref{sec:discussion} contains final discussion and conclusions.

\section{The 1D compass model and its solution}
\label{sec:Hamiltonian}

\subsection{From a frustrated magnet to spinless fermions}
\label{sec:fermions}

In this paper, we consider a frustrated 1D magnet with DM~interaction,
which relates to the magnetostriction and inverse DM (or spin-current)
mechanisms in a nonstoichiometric structure. To focus attention,
we assume that the 1D chain is along $x$ axis, i.e.,
$\hat{e}_{ij}=\hat{x}$, so the possible nonzero components of
$\vec{P}_i$ are $P^y_i$ and/or $P^z_i$ according to Eq.
(\ref{polarization}). In analogy to the 2D GCM \cite{Cin10},
we consider interactions which interpolate between the Ising model and
the frustrated interactions in the 1D compass model.
Therefore we introduce an arbitrary angle $\pm\theta/2$ relative to
$\sigma_l^x$ for odd/even bond, which defines new operators as linear
combinations of $\{\sigma_{l}^x,\sigma_{l}^y\}$ spin components on
even bonds,
\begin{eqnarray}
\tilde{\sigma}_{i}(\theta)&\equiv& \cos(\theta/2)\,\sigma_{i}^x
+\sin(\theta/2)\,\sigma_{i}^y,
\end{eqnarray}
and similar operators with angle $-\theta/2$ for odd bonds.
In such frustrated systems, the Ising-like interactions along odd
and even bonds have different strength and preferential easy-axes
within the ($\sigma^x,\sigma^y$) plane along the chain of $N$
sites.

The 1D GCM considered below is given by
\begin{eqnarray}
\cal{H}_{\rm GCM}&=& \sum_{i=1}^{N^\prime}\left\{
J_{o}\tilde{\sigma}_{2i-1}(\theta)\tilde{\sigma}_{2i}(\theta)
+J_{e}\tilde{\sigma}_{2i}(-\theta)\tilde{\sigma}_{2i+1}(-\theta)\right\}
\nonumber \\
&+&\sum_{i=1}^{N}\left\{ \vec{D}_{i}\cdot (\vec{\sigma}_{i}\times
\vec{\sigma}_{i+1} )+ \vec{E} \cdot \vec{P}_{i}
+ \vec{H} \cdot \vec{\sigma}_{i}\right\},
\label{Hamiltonian1}
\end{eqnarray}
where $J_o$ and $J_e$ denotes the coupling strength along odd and even
bonds, respectively. For convenience an even number of sites
$N$=$2 N^\prime$ is assumed. Here $\vec{E}$ is the electric field and
$\vec{H}$ is the external magnetic field, which contains the $g$-factor
$g$ and the Bohr magneton $\mu_{\rm B}$. $\vec{D}$ is the DM vector and
the interaction comes from a relativistic correction to the usual
superexchange that has a strength proportional to the spin-orbit
coupling constant. Without an inversion center on any bond,
the antisymmetric exchange is usually by one order of
magnitude smaller than the exchange interaction. An immediate question
is to what extent they affect the nonmagnetic phase. Note that the
first term in the Hamiltonian (\ref{Hamiltonian1}) interpolates between
the 1D Ising model ($\theta=0$) and the 1D compass model
($\theta=\pi/2$) \cite{Brz07,You08,You12}.
At an intermediate value of $\theta=\pi/3$ the interactions correspond
to $e_g$ orbitals --- such a model
was recently introduced for a 1D zigzag chain in an $(a,b)$ plane
\cite{You14}, and may be realized either in layered structures of
transition metal oxides \cite{Xiao11}, or in optical lattices
\cite{Simon11,Gsun}.

First, we consider below the model (\ref{Hamiltonian1}) in absence of an
external magnetic field. The role of magnetic field is explored in the
subsequent Sec. \ref{sec:HMF}. Without a magnetic field, the Hamiltonian
is invariant under a rotation in spin space around $z$-axis, and under
time reversal operator ${\cal T}=i\sigma_yK$, which reverses the sign of
all spin component operators. Here $K$ denotes the complex conjugation
operator and $\sigma_y$ is a Pauli matrix in spin space. As soon as
a DM interaction is introduced, the Hamiltonian is no longer invariant
with respect to a space inversion about a bond center. The DM
interaction always induces an electric
polarization according to Eq. (\ref{polarization}) that lies within
the rotation plane of the spins and is perpendicular to the magnetic
bond, and thus competes with the exchange energy. Here we presume that
the $\vec{D}$ vector is along the direction perpendicular to the plane,
i.e., $\vec{D}_{i}=D^z\hat{z}$, and originates from symmetry breaking
associated with the planar molecular structure that determines the
$(a,b)$ plane. In-plane components of the DM vector are assumed
negligible in comparison with the out-of-plane components.
By Eq. (\ref{polarization}) an electric field component $E^y$ along
the in-plane $y$-direction acts on the chiral polarization
$(\vec{\sigma}_{i}\times\vec{\sigma}_{i+1} )$ in the same way as the
$D^z$ component of the DM vector. Below we shall express this
dependence, for the considered geometry, by the variable $E$ defined as
\begin{equation}
E= D^z + \gamma E^y.
\label{E_parameter}
\end{equation}
Thereby the parity breaking field $E$ represents the $D^z$ component of the DM interaction
in addition controlled by the external electric field component  $E^y$.
Here $\gamma$ is the intrinsic material parameter that describes the
strength of the magneto-electric coupling of the system.
We note, that knowing the mapping in  Eq. (\ref{E_parameter}), which
is different for different materials, one can directly use all
our relations which feature the dependence on the field $E$.
Interestingly the control of DM interaction may also be extended
to oscillating electric fields which may allow for an
externally driven  rotation of spins.\cite{Chen14}.

The Hamiltonian (\ref{Hamiltonian1}) can be exactly diagonalized by
following the standard procedures. The Jordan-Wigner transformation
maps explicitly between spin operators and spinless fermion operators
by the following relations \cite{EBarouch70},
\begin{eqnarray}
\sigma _{j}^{+}& =&\exp \left[ i \pi \sum_{i=1}^{j-1}c_{i}^{\dagger }c_{i}
\right] c_{j}=\prod_{i=1}^{j-1}\sigma _{i}^{z}c_{j},  \notag \\
\sigma _{j}^{-}& =&\exp \left[ -i\pi \sum_{i=1}^{j-1}c_{i}^{\dagger }c_{i}
\right] c_{j}^{\dagger }=\prod_{i=1}^{j-1}\sigma
_{i}^{z}c_{j}^{\dagger },
\notag \\
\sigma _{j}^{z}& =&1-2c_{j}^{\dagger }c_{j},
\end{eqnarray}
where $c_{j}$ and $c_{j}^{\dagger }$ are annihilation and creation
operators of spinless fermions at site $j$, which obey the standard
anticommutation relations: $\{c_i, c_j\}=0$,
$\{c_i^{\dagger},c_j\}=\delta_{ij}$.
Consequently, we have a free-fermion Hamiltonian:
\begin{eqnarray}
\hat{H}_{E}\!&=&\!
\sum_{i} \left[J_{o}e^{ i\theta} c_{2i-1}^{\dagger} c_{2i}^{\dagger}
  + (J_{o}-2 i E) c_{2i-1}^{\dagger} c_{2i} \right. \nonumber \\
\!&+&\!\left.J_{e}  e^{-i\theta} c_{2i}^{\dagger} c_{2i+1}^{\dagger}
+ (J_{e}-2 i E) c_{2i}^{\dagger} c_{2i+1}+{\rm H.c.}\right].
\label{Hamiltonian2}
\end{eqnarray}
The spinless fermion Hamiltonian is equivalent to the 1D mean-field
model for a triplet superconductor, with inhomogeneous nearest neighbor
hopping and condensate amplitudes \cite{Kitaev01}.

\subsection{Quasiparticles at finite electric field and $\vec{H}=0$}
\label{sec:quasi}

The above Hamiltonian can be diagonalized; to this end we introduce
the discrete Fourier transformation of the fermionic operators,
\begin{eqnarray}
c_{2j-1}=\!\frac{1}{\sqrt{N'}}\sum_{k}e^{-ik j}a_{k},\hskip .2cm
c_{2j}=\!\frac{1}{\sqrt{N'}}\sum_{k}e^{-ik j}b_{k}, \quad
\end{eqnarray}
with the discrete momenta given as follows,
\begin{eqnarray}
k=\frac{n\pi}{ N^\prime  }, \quad n= -(N^\prime\!-1), -(N^\prime\!-3),
\ldots, (N^\prime\! -1). \quad
\end{eqnarray}
The Hamiltonian takes the following form which is suitable to
introduce the Bogoliubov transformation,
\begin{eqnarray}
\hat{H}_{E}=\! \sum_{k} \left[ B_k a_{k}^{\dagger}b_{-k}^{\dagger}
+\! A_k a_{k}^{\dagger} b_{k}-\! A_k^* a_{k}b_{k}^{\dagger}
-\!B_k^* a_{k}b_{-k} \right].\quad \quad
\label{Hamiltonian5}
\end{eqnarray}
Here
\begin{eqnarray}
A_k&=& (J_{o}-2iE)+ (J_{e}+ 2iE) e^{ik},  \nonumber \\
B_k&=& J_oe^{i\theta}-J_e e^{i(k-\theta)}.
\end{eqnarray}
To diagonalize the Hamiltonian Eq. (\ref{Hamiltonian5}), we rewrite
it in the Bogoliubov-de Gennes (BdG) form,
\begin{eqnarray}
\hat{H}_{\rm 0} &=& \sum_{k}
\Gamma_k^{\dagger}
\hat{M}_k
\Gamma_k, \label{FT2}
\end{eqnarray}
where
\begin{eqnarray}
\hat{M}_k=\frac{1}{2}\left(\begin{array}{cccc}
0 & 0 &  R_k+S_k   &  P_k+Q_k  \\
0 & 0 & P_k- Q_k   &   R_k-S_k  \\
R_k^*+S_k^*  &  P_k^*-Q_k^*   & 0 & 0 \\
P_k^*+Q_k^*   & R_k^*-S_k^*   & 0 &0
\end{array}\right) \nonumber \\ \label{Mk}
\end{eqnarray}
and $\Gamma_k^{\dagger} =(a_{k}^{\dagger},a_{-k},b_k^{\dagger},b_{-k})$.
Here we have defined:
\begin{eqnarray}
P_k&=&-i (J_e e^{ik}+J_o)\sin\theta, \nonumber \\
Q_k&=& (J_e e^{ik}-J_o)\cos\theta,  \nonumber \\
S_k &=&J_o+J_e e^{ik}, \nonumber \\
R_k &=& 2 i E ( e^{ik} -1 ). \label{terms}
\end{eqnarray}
Within the Majorana representation, the BdG Hamiltonian (\ref{Mk}) acts
in an enlarged expanded Nambu-spinor space, namely the tensor product of
the physical space $\textbf{C}^{2N}$ with an extra degree of freedom
$\textbf{C}^{2}$, which we call the "particle-hole space"
\cite{Altland}. This structure has an emergent particle-hole symmetry
(PHS) ${\cal C}=\tau_x K$, namely, $\{\hat{M}_k, {\cal C}\}=0$. Here,
$\tau_x$ is a Pauli matrix acting in the Nambu space. Noting that both
time-reversal operator ${\cal T}$ and particle-hole transformation
${\cal C}$ are anti-unitary operators, satisfying $[\hat{H},{\cal T}]=0$,
$\{\hat{H},{\cal C}\}=0$. As a consequence, two copies of the actual
excitation spectrum, a particle and a hole copy, emerge simultaneously
\cite{Budich}.

A unitary transformation $\hat{U}_k$ can transform the Hermitian matrix
(\ref{Mk}) into a diagonal form,
\begin{eqnarray}
\hat{\Upsilon}_k=\hat{U}_k \hat{M}_k \hat{U}_k^{\dagger}.
\end{eqnarray}
The quasiparticle (QP) operators,
$\{\gamma_{k,1}^{\dagger},\gamma_{k,2}^{\dagger},\gamma_{k,3}^{\dagger},
\gamma_{k,4}^{\dagger}\}$, are connected with
$\{a_k^{\dagger},a_{-k}^{},b_k^{\dagger},b_{-k}^{}\}$ through the
following relation,
\begin{eqnarray}
\left(
\begin{array}{c}
\gamma_{k,1}^{\dagger} \\
\gamma_{k,4}^{\dagger}  \\
\gamma_{k,2}^{ \dagger}\\
\gamma_{k,3}^{\dagger}
\end{array}
\right)=\hat{U}_{k} \left(
\begin{array}{c}
a_k^{\dagger}  \\
a_{-k}   \\
b_k^{\dagger}   \\
b_{-k}
\end{array}%
\right). \label{eq:2DXXZ_RDM}
\end{eqnarray}%
After diagonalization, the eigenspectra $\varepsilon_{k,j}$
($j=1,\cdots, 4$) are readily obtained:
\begin{eqnarray}
\varepsilon_{k,1(2)}=
-\frac{1}{2}\sqrt{\varsigma_k\pm \sqrt{\varsigma_k^2-\tau_k^2}},\\
\varepsilon_{k,3(4)}=
 \frac{1}{2}\sqrt{\varsigma_k\mp \sqrt{\varsigma_k^2-\tau_k^2}},
\label{excitationspectrum}
\end{eqnarray}
where
\begin{eqnarray}
\varsigma_k&=&\vert P_k \vert^2 + \vert Q_k \vert^2 + \vert R_k
\vert^2+ \vert S_k \vert^2, \nonumber \\
\tau_k&=&\vert  P_k^2 - Q_k^2 - R_k^2 + S_k^2 \vert.
\end{eqnarray}

\begin{figure}[t!]
\includegraphics[width=8cm]{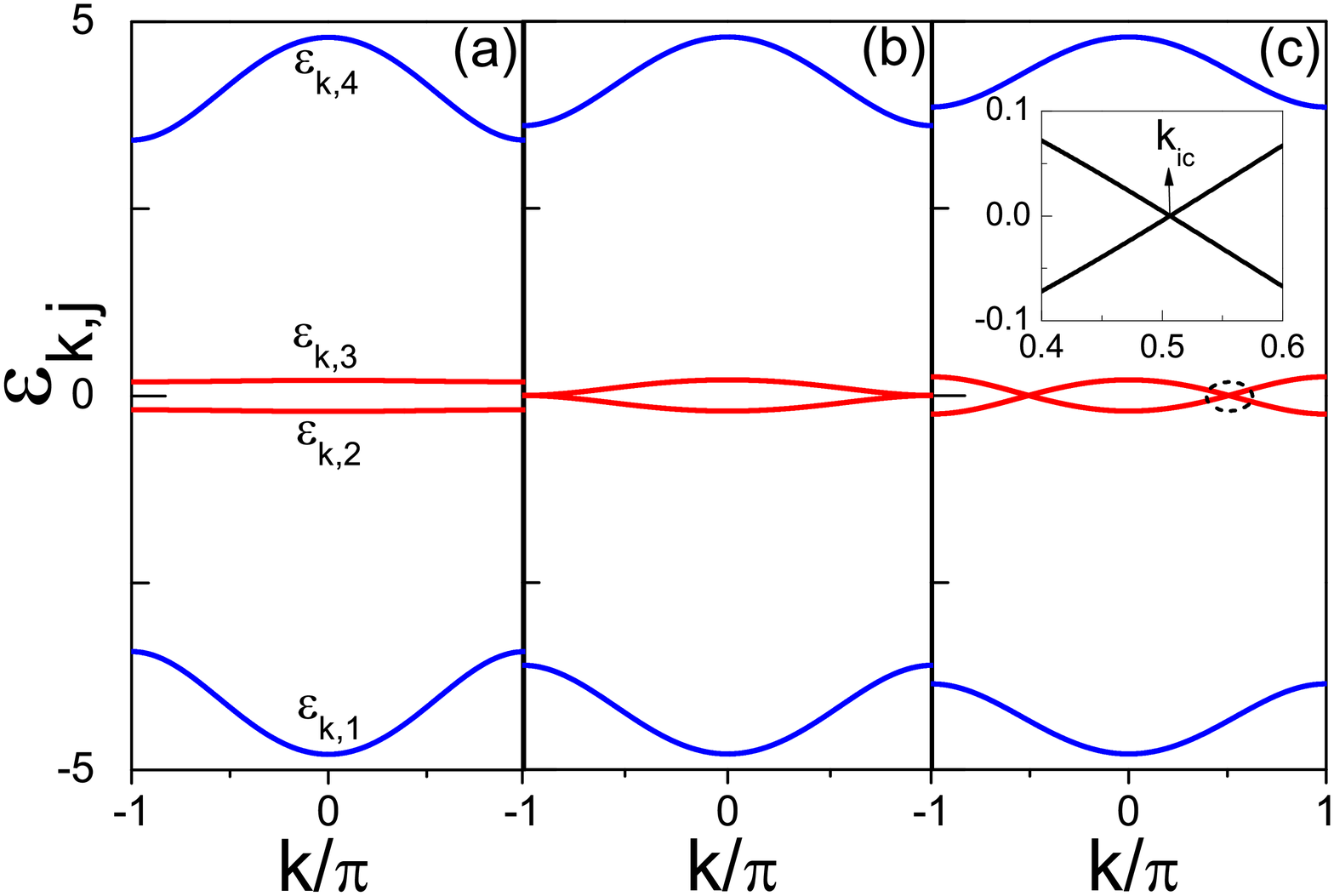}
\caption{(Color online)
The energy spectra $\varepsilon_{k,j}$ ($j=1,\cdots,4$) for
increasing electric field $E$:
(a) $E=0.3$,
(b) $E=0.5$, and
(c) $E=0.7$.
Inset in (c) is the amplification of the dashed circle below.
Parameters are as follows: $J_o=1$, $J_e=4$, $\theta=\pi/3$, $H=0$. }
\label{spectrum-E}
\end{figure}

The eigenenergies are labeled sequentially from the bottom to the top as
$\varepsilon_{k,1},\cdots,\varepsilon_{k,4}$, see Fig. \ref{spectrum-E}.
Note that the spectra of mode $k=0$ are independent of $E$, in contrast
to the other modes. We can make this peculiarity to trace the order of
spectra. Instantly we obtain the diagonal form of the Hamiltonian,
\begin{eqnarray}
\hat{H}_{\rm 0}=\sum_{k}\sum_{j=1}^{4}  \varepsilon_{k,j}
\gamma_{k,j}^{\dagger}\gamma_{k,j}^{}.
\label{diagonalform}
\end{eqnarray}
One finds that the spectra are symmetric with respect to energy
$\varepsilon=0$ and the $k\leftrightarrow -k$ transformation; see the
QP bands in Fig. \ref{spectrum-E}. The positive spectra
correspond to the electron excitations while the negative ones are the
corresponding hole excitations. As seen in Fig. \ref{spectrum-E}(a),
the upper two branches of the spectra, $\varepsilon_{k,3}$ and
$\varepsilon_{k,4}$, are always positive for $E=0.3$. The PHS implies
here that $\gamma_{k,4}^{\dagger}$=$\gamma_{-k,1}$,
$\gamma_{k,3}^{\dagger}$=$\gamma_{-k,2}$.
Accordingly, the gap is determined by the absolute value of the
difference between the second and third energy branches,
\begin{equation}
\Delta=\min_{k}\vert  \varepsilon_{k,2}- \varepsilon_{-k,3}\vert.
\end{equation}

With the increase of $E$, $\varepsilon_{ \pi,3}$ bends down and
$\varepsilon_{\pi,2}$ moves upwards at $k\approx \pm\pi$. Finally,
$\varepsilon_{\pi,3}$ touches $\varepsilon_{\pi,2}$ at $E=0.5$ and
$k=\pm\pi$, i.e., $\Delta=0$; cf. Fig.\ref{spectrum-E}(b). The condition
for the gap closing requires $\tau_\pi=0$, which gives rise to
\begin{equation}
E_c \equiv  \frac12\sqrt{J_o J_e}\, \vert \cos \theta \vert\,.
\label{conditionforcriticalh2}
\end{equation}
Further increase of $E$ leads to the bands inversion;
$\varepsilon_{\pi,2}$ and $\varepsilon_{\pi,3}$ cross at two
generally incommensurate and symmetric momenta $\pm k_{\rm ic}$,
which is given by
\begin{eqnarray}
k_{\rm ic}=\arccos\left(1-\frac{J_o J_e \cos^2 \theta}{2 E^2}\right).
\end{eqnarray}
One finds that in Fig. \ref{spectrum-E}(c) the energies in the
upper second band can be negative,
\begin{eqnarray}
\varepsilon_{k,3}&\le& 0, \quad {\rm for}\;\vert k\vert\ge k_{\rm ic}.
\end{eqnarray}

\begin{figure}[t!]
\includegraphics[width=8cm]{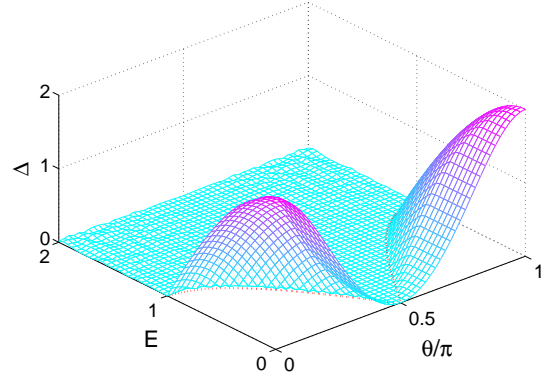}
\caption{(Color online)
The gap $\Delta$ as a function of $E$ and $\theta$.
The dotted lines are obtained from Eq. (\ref{conditionforcriticalh2}).
They separate a gapless chiral phase from two gapful, canted
N\'{e}el phases, that are separated by a quantum critical point
at $\theta=\pi/2$ .
Parameters are as follows: $J_o$=1, $J_e$=4. }
\label{Gap2}
\end{figure}

The ground state of any fermion system follows the total filling of
the Fermi-Dirac statistics, and the lowest energy is obtained when all
the QP states with negative energies are filled by fermions. More
precisely, in the thermodynamic limit ($N\to\infty$), the ground state
of the system, $|\Phi_0\rangle$,
corresponds to the configuration with chemical potential $\mu=0$, where
all the states with $\varepsilon_{k,j}<0$ are occupied and the ones
with $\varepsilon_{k,j}\ge 0$ are empty. By means of the corresponding
occupation numbers
\begin{equation}
n_{k,j}=\langle\Phi_0\vert\gamma_{k,j}^{\dagger}\gamma_{k,j}^{}
\vert\Phi_0\rangle = \left\{
  \begin{array}{l l}
    0 & \quad {\rm for}\;\varepsilon_{k,j} \ge 0,\\
    1 & \quad{\rm for}\;\varepsilon_{k,j}<0.
  \end{array} \right.
\end{equation}
One recognizes that in the present case of a symmetric QP spectrum,
the ground state energy may be expressed as
\begin{eqnarray}
E_0 = -\frac{1}{2} \sum_{k} \sum_{j=1}^4\vert \varepsilon_{k,j} \vert.
\label{E0expression}
\end{eqnarray}
The advantage of the result given by Eq. (\ref{E0expression}) is that
it is independent of the signs of $J_o$ and $J_e$ which can be verified
by transformation $\sigma_{2i-1}^x\to -\sigma_{2i-1}^x$ and
$\sigma_{2i}^y \to -\sigma_{2i}^y$.

As observed in Fig. \ref{Gap2}, the gap $\Delta$ diminishes after $E$
exceeds the critical value $E_c$ for a fixed angle $\theta$; $E_c$ is
symmetric with respect to $\theta=\pi/2$, and decreases with $\theta$.
As $E$ approaches $E_c$ (\ref{conditionforcriticalh2}) from below, the
size dependence of the gap, $\Delta\sim L^{-z}$, defines the dynamic
exponent $z$. Expanding the gap around the critical line $E_c$ from
lower threshold, i.e., at $\tau_k\to 0$,
\begin{eqnarray}
\Delta &\sim&  \frac{\tau_k}{\sqrt{2\varsigma_k}}
 \sim \frac{8(E_c^2-E^2)}{\sqrt{J_o^2+J_e^2}}.
\end{eqnarray}
The relativistic spectra at $k_{\rm ic}$ imply a dynamical exponent
$z=1$ for $\theta \neq \pi/2$; see inset in Fig. \ref{spectrum-E}(c).
The linear dispersion law guarantees that the density of low-energy
states in the anisotropic chain remains finite instead of leading to the
square-root divergence typical for isotropic spin chains
\cite{Zhitomirsky04}. In contrast, the point $\theta=\pi/2$ at $E=0$
is a multicritical point with an emergent $\mathbb{Z}_2$ symmetry and
the spectra vanish quadratically at $\pm \pi$ as a result of the
confluence of two Dirac points, corresponding to a dynamical exponent
$z = 2$ \cite{Niu12}.

\subsection{Correlation functions}
\label{sec:corr}

In order to characterize the QPTs, we studied
the nearest neighbor spin correlation function $C_e^{\alpha}$ ($C_o^{\alpha}$)
on even (odd)  bonds  defined by
\begin{eqnarray}
C^{\alpha}_{l}&=&-\frac{2}{N}\sum_{i=1}^{N/2}\langle
\sigma_{i}^\alpha
\sigma_{i+l}^\alpha \rangle,
\end{eqnarray}
where $l$=1(-1) and
the superscript $\alpha=x,y,z$ denotes the cartesian component,
and chirality correlation function,
\begin{eqnarray}
X_{l}^{\alpha}&=&-\frac{2}{N}\sum_{i=1}^{N}\langle
\vec{\alpha} \cdot
(\vec{\sigma}_{i}\times \vec{\sigma}_{i+l} )\rangle,
\label{chir}
\end{eqnarray}
where $\vec{\alpha}$ denotes the unit vector in the direction of a
cartesian component $\alpha$.
The chirality $X_{l}^{\alpha}$
(\ref{chir}) will exhibit a sign change under the parity operation but
stay invariant under the time-reversal operation.
Finally, we introduce the nonlocal string order parameter,
\begin{eqnarray}
 O_{s}^{\alpha} = \langle
S^{\alpha}_{4k} S^{\alpha}_{4k+1}S^{\alpha}_{4k+2}S^{\alpha}_{4k+3}
\cdots S^{\alpha}_{4n}
S^{\alpha}_{4n+1}S^{\alpha}_{4n+2}S^{\alpha}_{4n+3} \rangle.
\nonumber \\
\end{eqnarray}
The string order parameters can
be understood as an extension of the two-site
correlation function, that essentially captures the hidden topological order in low-dimensional quantum systems \cite{Nijs}.
They provides supplementary description for those quantum phases that are not amenable to a charterization through local order parameters.
It was demonstrated that the nonlocal
string order was directly measured in
one-dimensional bosonic Mott insulator \cite{Endres}.
These correlation functions can be calculated from the
two-point correlation functions that can be obtained as determinants
as a result of Wick's theorem \cite{EBarouch70,Osb02},
\begin{eqnarray}
\langle \sigma_0^x \sigma_r^x \rangle &=& \left \vert
\begin{array}{c c c c}
G_{-1} & G_{-2} & \cdot & G_{-r} \\
G_{0} & G_{-1} & \cdot & G_{-r+1} \\
\vdots & \vdots & \ddots &\vdots  \\
G_{r-2} & G_{r-3} & \cdot & G_{-1}
\end{array} \right \vert, \\
\langle \sigma_0^y \sigma_r^y \rangle &=& \left \vert
\begin{array}{c c c c}
G_{1} & G_{0} & \cdot & G_{-r+2} \\
G_{2} & G_{1} & \cdot & G_{-r+3} \\
\vdots & \vdots & \ddots &\vdots  \\
G_{r} & G_{r-1} & \cdot & G_{1}
\end{array} \right \vert, \\
\langle \sigma_0^z\sigma_r^z \rangle &=& 4\langle \sigma_0^z
\rangle \langle \sigma_r^z \rangle
- G_r G_{-r},
\end{eqnarray}
where
\begin{eqnarray}
G_r=\langle \sigma_0^y \sigma_r^x \rangle.
\end{eqnarray}

\begin{figure}[t!]
\includegraphics[width=8cm]{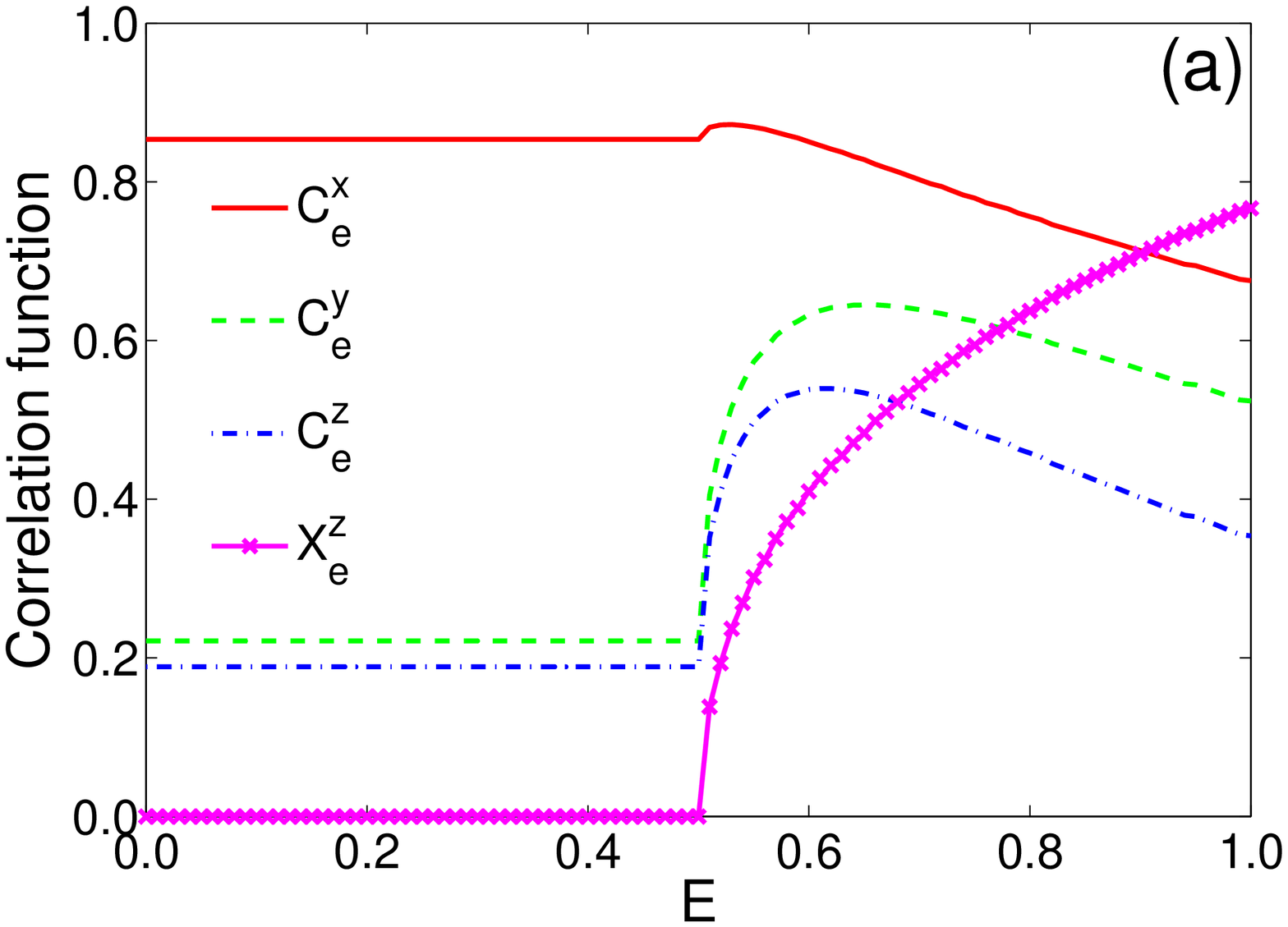}
\includegraphics[width=8cm]{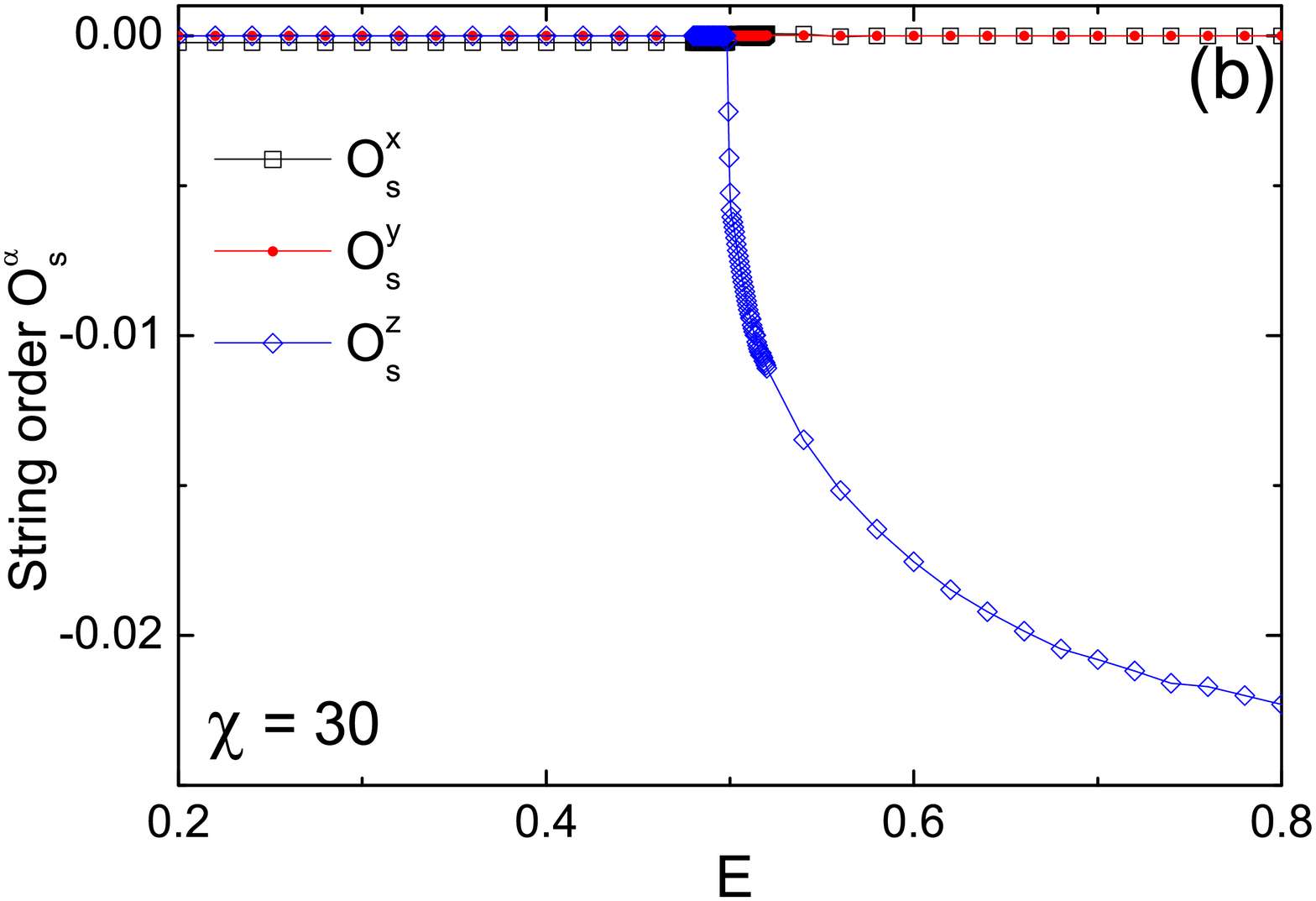}
\caption{(Color online) Evolution of the ground state for increasing
electric field:
(a) the nearest-neighbor correlations $C^\alpha$ and chirality
$X^\alpha$ on even bonds;
(b) the string order parameters $O^{\alpha}_s$.
The bond dimension is set as $\chi=30$.
Parameters are as follows: $n - k  = 200$, $J_o=1$, $J_e=4$,
$\theta=\pi/3$, and $H$=0. }
\label{Orderparameters-h=0}
\end{figure}

When $E=0$, the ground state is marked by the finite nearest neighbor
correlation functions, among which $x$-components $\{C_l^x\}$ dominate
for $\theta<\pi/2$, implying that the adjacent spins are antiparallel
and aligned with a canted angle with respect to the $x$ axis. In other
words, the ground state of the GCM is a canted N\'eel (CN) phase for
$\theta<\pi/2$. The predicted spin-wave spectrum can be compared with
the results obtained from inelastic neutron scattering measurements
\cite{Jeong12,Matsuda}. Conversely, the $z$-component chirality
$X_l^{z}$ completely vanishes. With increase of $E$, $C_l^{x}$,
$C_l^{y}$ and $C_l^{z}$ remain unchanged until reaching a threshold
value $E_c(\theta)$, as shown in Fig. \ref{Orderparameters-h=0}(a).
Simultaneously, the chiral order $X^z$ starts to grow and saturates as
$E\to\infty$. The DM interaction or the electric field induces spins
to be cycloidally oriented in the $(\sigma^x,\sigma^y)$ plane.

Effects of bond alternation and the DM interaction on the
zero-temperature phase diagram of the Ising model has been studied in
terms of an infinite time-evolving block decimation (iTEBD) algorithm
\cite{Liu13}. The iTEBD method allows one to solve for the ground state
properties of a 1D translationally invariant spin system of infinite
length. One of the main controlling factors under this strategy is the
bond dimension $\chi$, i.e., the cut-off dimension of Schmidt
coefficients during singular value decomposition process \cite{Vidal07}:
\begin{eqnarray}
|\Psi\rangle=
\sum_{i=1}^{\chi}\vert\phi_i^L\rangle\Lambda_i\vert\phi_i^R\rangle,
\end{eqnarray}
where $\vert\phi_i^L\rangle$ and $\vert\phi_i^R\rangle$ represent the
orthonormal bases of the subsystem to the left and right halves of the
broken bond, and $\Lambda$ is a diagonal matrix. The iTEBD algorithm can
not only reveal ground state energy, excitations, local correlation
functions, but also can conveniently evaluate nonlocal correlations,
like quantum entanglement, which are not easy to obtain by other methods.

\begin{figure}[t!]
\includegraphics[width=8cm]{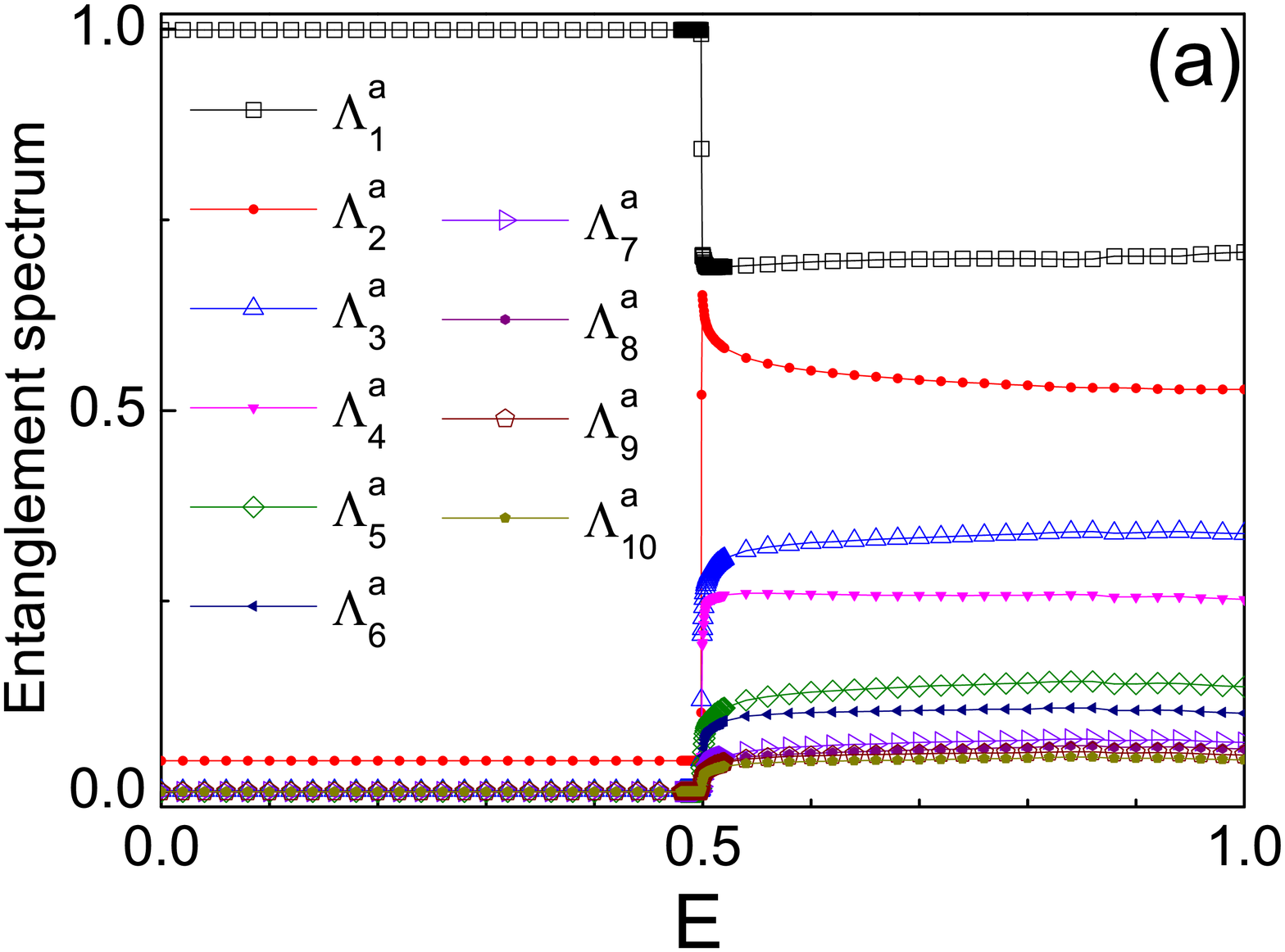}
\includegraphics[width=8cm]{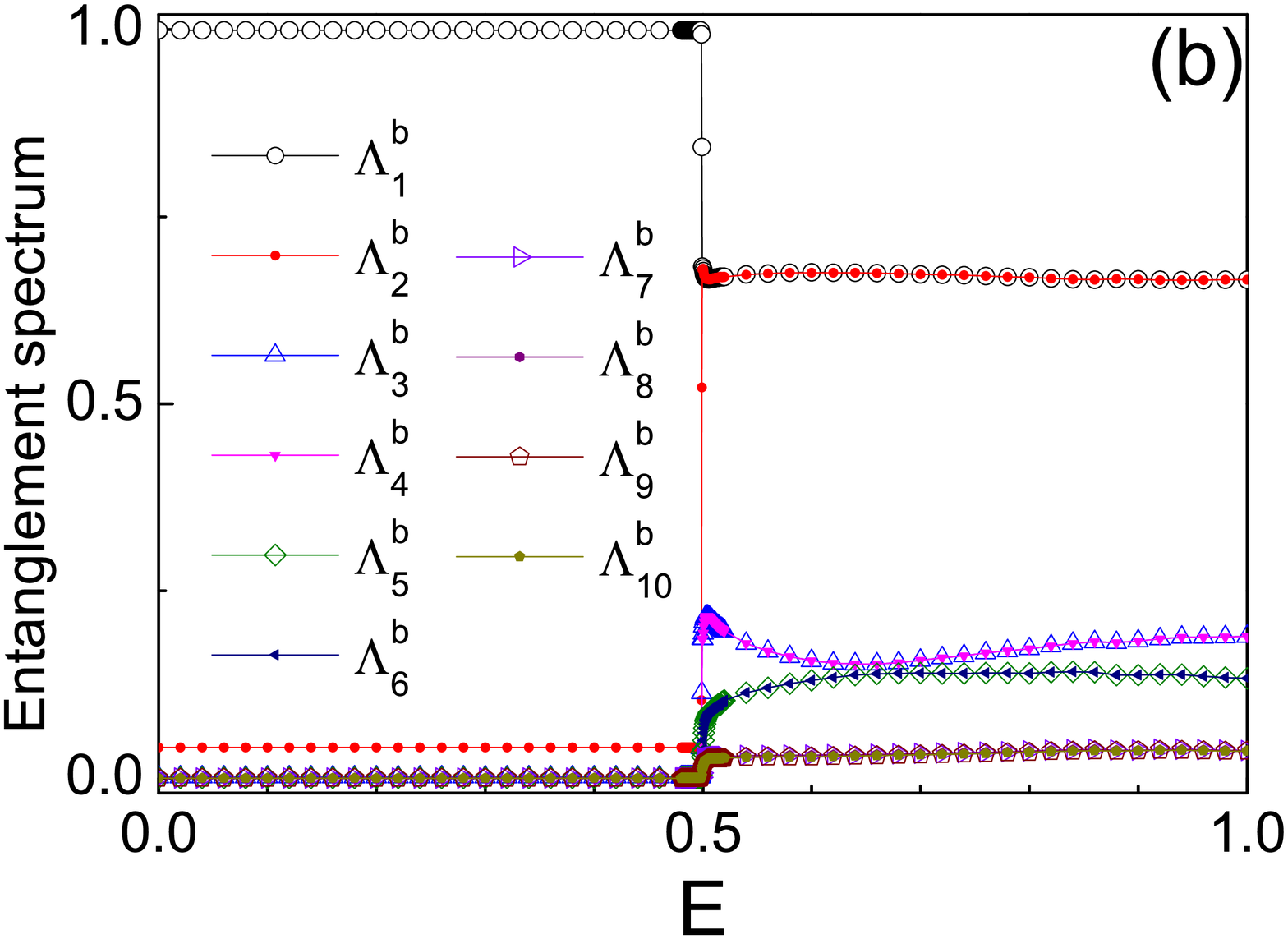}
\caption{(Color online) The normalized entanglement spectrum for
increasing electric field $E$ by cutting a single:
(a) odd bond, and
(b) even bond.
Parameters are as follows: $J_o$=1, $J_e$=4, $\theta=\pi/3$, and $H$=0. }
 \label{ES1}
\end{figure}

Quantum entanglement was originally introduced to exhibit nonlocal
correlations in a quantum system \cite{Amico,Eisert,You1112,Ole12}.
Considering a $d$-dimensional block composed by $L^d$ contiguous spins,
embedded in an infinite system, the von Neumann entropy between the
block and the rest of the system is given by
\begin{eqnarray}
{\cal S}_{ L}= -\textrm{Tr} \rho_{L} \log_2 \rho_{L},
\end{eqnarray}
where $\rho_{L}$ is the reduced density matrix for the $L^d$-site block.
A celebrated boundary law is satisfied for
\begin{eqnarray}
{\cal S}_{ L} \sim L^{d-1}
\label{SL}
\end{eqnarray}
in a gapped $d$-dimensional system due to a short-range correlation
length \cite{Gottesman,Calabrese-2004}, while a logarithmic additive
term in gapless regimes becomes dominated by the form
\cite{Barthel,Gioev,Klich,Song11,Song12}
\begin{eqnarray}
S_L \sim \frac{c}{3} L^{d-1} \log_2 L, \label{SL-L}
\end{eqnarray}
where $c$ is central charge. Moreover, the study of the entanglement
spectrum, i.e., the eigenvalues $\xi_i$ of entanglement Hamiltonian
$H_L$ resulting from $\rho_L=e^{-H_{\rm L}}$ \cite{Haldane}, has been
recognized that the universal part of entanglement spectrum reveals an
intricate connection between a bulk property and edge physics
\cite{Swingle2012}.

An alternative way to look at the dimerization of a chain is via the
study of entanglement entropy of weak and strong bonds \cite{Sirker08}.
In Fig. \ref{ES1}, the normalized entanglement spectra of the
half-infinite chain, obtained by dividing the chain into two
half-infinite chains, are shown as functions of $E$. Since Eq.
(\ref{Hamiltonian1}) is a two-period system, we have two entanglement
spectra ($\Lambda^a$) and ($\Lambda^b$) by cutting odd or even bonds.
They are not equivalent ---
we find that the entanglement spectra $\Lambda^b$ are doubly degenerate
in chiral phases, in contrast to $\Lambda^a$. The exact two-fold
degeneracy in the entire entanglement spectrum is protected by the
space inversion (parity) symmetry of "odd parity"-chiral state
\cite{Pollmann3}, and this implies the existence of a nonlocal string
order parameter \cite{Pollmann}. The iTEBD calculation reveals that
nonlocal correlation $O_s^z$ arises for $E>E_c$, observed in Fig.
\ref{Orderparameters-h=0}(b).
The bipartite entanglement between two
half-infinite chains can be directly read out through
\begin{eqnarray}
S_{2i-1,2i}&=&-{\rm Tr} [(\Lambda^a)^2 \log_2(\Lambda^a)^2], \\
S_{2i,2i+1}&=&-{\rm Tr} [(\Lambda^b)^2 \log_2(\Lambda^b)^2].
\end{eqnarray}
The bipartite entanglement on even bonds is larger than that on odd
bonds, and both of them exhibit a singularity at criticality. In the
iTEBD calculation the divergence of $S_{\rm vN}$ at the critical point
is argued to scale with bond dimension $\chi$ like \cite{Pollmann2},
\begin{eqnarray}
S_{\rm vN} \sim\frac{1}{\sqrt{12/c}+1}\ln \chi.
\end{eqnarray}

\begin{figure}[t!]
\includegraphics[width=8cm]{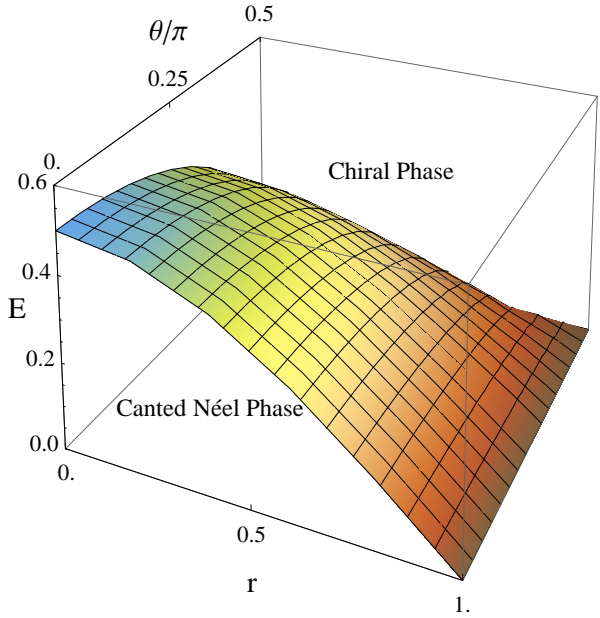}
\caption{(Color online) Ground state phase diagram of Eq.
(\ref{Hamiltonian1}), at zero magnetic field, as function of the
dimerization parameter Eq. (\ref{r}), the electric field $E$ and
angle $\theta$.
}
\label{PD1}
\end{figure}

A QPT is indicated to exist in the zero-temperature phase diagram by
nonanalyticity of order parameters with the controlling parameter.
Thus, one finds a QPT from the gapped N\'eel phase to the gapless
chiral phase at the critical point $E_c$. The three-dimensional phase
diagram as functions of varying angle $\theta$, dimerization parameter,
\begin{equation}
r\equiv\frac{J_e-J_o}{J_e+J_o},
\label{r}
\end{equation}
and electric field $E$, is presented in Fig. \ref{PD1}. The transition
from the CN phase to the chiral phase occurs at the critical value of
the electric field $E_c$ given by Eq. (\ref{conditionforcriticalh2}).
The CN phase is stable in a finite range of $0<E<E_c$, except for the
limit of decoupled dimers on even bonds ($r=1$), or the value of
$\theta=\pi/2$, where the chiral phase exists at any  electric field
strength $E$.

Until now we have focused on the role played by the DM interaction and
the applied electric field in the model. It is of interest now to ask
how the above scenario is modified by the additional effect of finite
magnetic field and we explore this problem in the following Section.

\section{ Generalized compass model \\ in a homogenous magnetic field }
\label{sec:HMF}

\begin{figure}[t!]
\includegraphics[width=8cm]{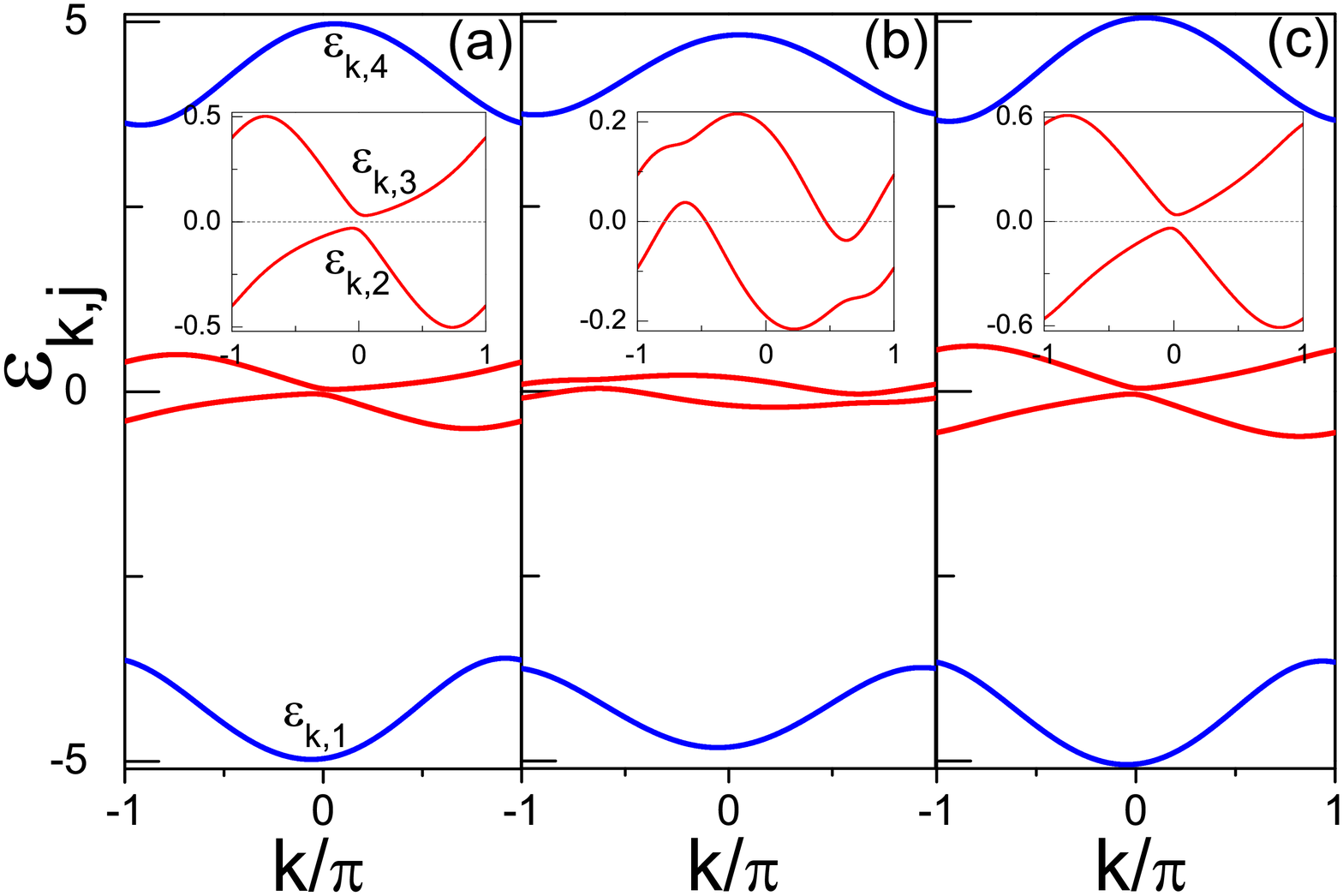}
\caption{(Color online)
The energy spectra $\varepsilon_{k,j}$ ($j=1,\cdots,4$) for:
(a) the CN-phase at $P_1$ ($E=0.3$, $H=1.8$),
(b) the chiral phase at $P_2$ ($E=0.6$, $H=0.6$), and
(c) the polarized phase at $P_3$ ($E=0.2$, $H=2.2$).
Insets show the amplification of the corresponding two
central branches. $\varepsilon_{k,3}$ is
negative for $k \in$ $(0.4655 \pi, 0.7871\pi)$ at $P_2$.
A horizonal guiding line marks the position of chemical potential,
i.e., $\mu=0$.
Parameters are as follows: $J_o=1$, $J_e=4$, $\theta=\pi/3$. }
\label{spectrum-k}
\end{figure}

Here we study the effect of a homogenous magnetic field and the
associated MEEs. We consider the case where the magnetic field is
oriented \textit{perpendicular to the easy-plane of the spins},
i.e., $\vec{H}=H\hat{z}$. Subsequently, in Nambu representation, the
Hamiltonian matrix is modified in the following way,
\begin{equation}
\hat{M}_k\to\hat{M}^\prime_k=\hat{M}_k-H{\mathbb I}_2\otimes\sigma^z,
\label{Hamiltonian6}
\end{equation}
where ${\mathbb I}_2$ is a $(2\times 2)$ unity matrix. The directional
Zeeman splitting perpendicular to the $(x,y)$ plane lifts the Kramers
degeneracy, and makes the expression for the ground state energy rather
involved, which will not be shown here. The analytical solution of
Hamiltonian (\ref{Hamiltonian6}) along the path $E=0$ had been
scrutinized recently, and an order-disorder QPT induced by the magnetic
field was recognized \cite{You14}. Such criticality is suited at
momentum $k=0$. Generally, the Hamiltonian (\ref{Hamiltonian1}) breaks
the space inversion symmetry of spin chain when electric field $E$ is
applied, while $H$ field breaks the time reversal symmetry (TRS).
Figure \ref{spectrum-k} shows the energy spectra for three typical
values of $E$ and $H$. The joint breaking of TRS and parity symmetry
leads to the asymmetry of the bands with respect to $k=0$.

\begin{figure}[t!]
\includegraphics[width=8cm]{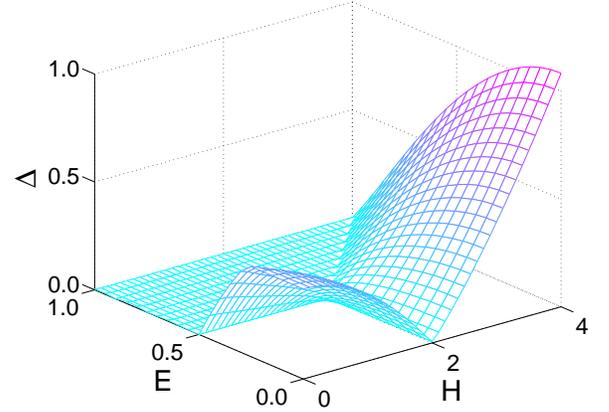}
\caption{(Color online)
The excitation gap $\Delta$ as functions of electric $E$ and magnetic
$H$ field. One recognizes the gapped CN (polarized) phase at small
(large) $H$, respectively, and the gapless chiral phase at large $E$.
Parameters are as follows: $J_o=1$, $J_e=4$, and $\theta=\pi/3$.
}
\label{Gap-D-h}
\end{figure}

The BdG band structure in the absence of further symmetries still
preserves the antiunitary PHS. The PHS of the BdG Hamiltonian is
\begin{eqnarray}
{\cal C} \hat{M}_{k} {\cal C} =-\hat{M}_{-k}.
\end{eqnarray}
Hence, we have
\begin{equation}
 \varepsilon_{k,1}=-\varepsilon_{-k,4},\;\;
\varepsilon_{k,2}=-\varepsilon_{-k,3}.
\end{equation}
Note that $k=0$ and $k=\pm\pi$ are special points; the latter are called
"time-reversal-invariant" points, since they are mapped onto
themselves. At $E=0.6$ and $H=0.6$, we can see from Fig.
\ref{spectrum-k}(b) that the energy spectrum $\varepsilon_{k,3}$ is not
positive for all $k$ values in the Brillouin zone, as this band crosses
from positive to negative values at some intermediate value of $k$. The
appearance of hole and electron pockets generate four Fermi points,
\textit{and is the key feature of the chiral phase at finite $H$-field.}
A scrutiny of gap for typical parameters in Fig. \ref{Gap-D-h}
incorporated Lee-Yang zeros \cite{YangLee,Wei2012,Liu2012} sketches
three different phases. Two of them are gapped in the excitation
spectrum while the third one is gapless.

\begin{figure}[t!]
\includegraphics[width=8cm]{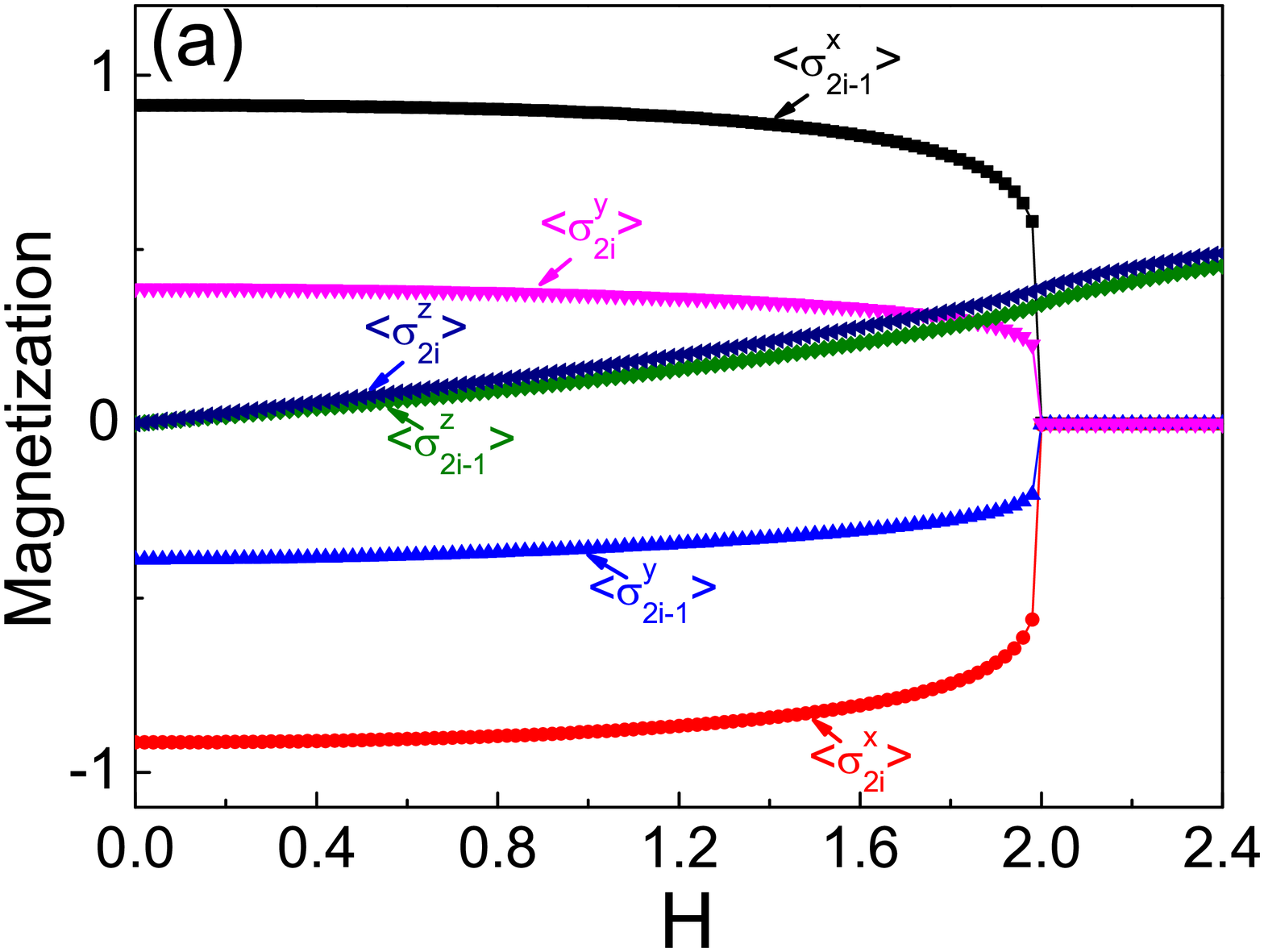}
\includegraphics[width=8cm]{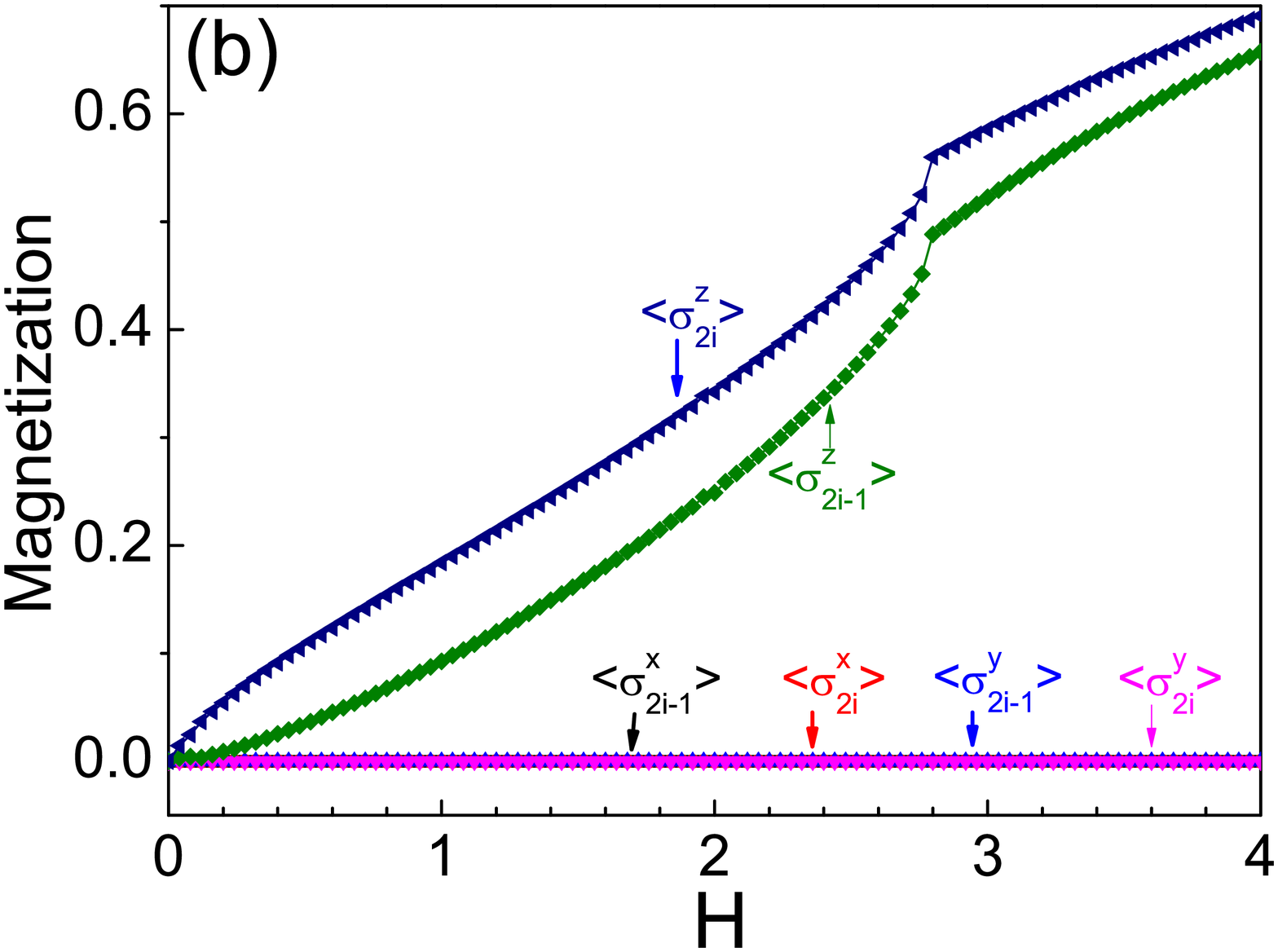}
\caption{(Color online) Evolution of magnetization components with
magnetic field $H$ for (a) $E$=0.3 and (b)$E$= 0.7.
Parameters are as follows: $J_o=1$, $J_e=4$, $\theta=\pi/3$. }
\label{Magnetization-h}
\end{figure}

The spin chirality $(\vec{\sigma}_i\times\vec{\sigma}_{i+1})$ is
perpendicular to the spin-spiral plane and is odd under spatial
reflections. The magnetic field $\propto H$ induces a tilt of
spin-spiral plane. Both the time-reversal and parity symmetries are
broken in this spin-spiral phase. As a result, it exhibits a MEE, and
thus leads to directional change of polarizations $P$ and staggered
magnetic moments, which can be measured by nuclear magnetic resonance (NMR) and muon spectroscopy ($\mu$SR)
\cite{Hassanieh}.

The corresponding components of magnetization are shown in Fig.
\ref{Magnetization-h} for increasing $H$. The magnetization is found to
be almost independent of $E$ as long as the system is within a given
magnetic phase, but discontinuous changes of magnetization occur at
phase transitions. At $E=0.3$, an abrupt change of each
$\sigma_i^{\alpha}$ component occurs at $H=2.0$, and only the
$z$-component of the magnetization survives for $H>2.0$, see Fig.
\ref{Magnetization-h}(a). A nonzero chirality $X_l^z$ shows up when
$H>2.0$. At $E=0.7$, the $x$-th and $y$-th magnetization components
completely disappear regardless of the value of $H$, see Fig.
\ref{Magnetization-h}(b). Only the $z$-th magnetization component is
monotonously enhanced by increasing magnetic field, and a discontinuity
occurs at $H=2.8$. The sharp downturn of the magnetization below the
critical field indicates a competition with the chiral order parameter
of the chiral phase, and the chirality $X_{l}^{z}$ decreases as $H$
grows and a kink also arises at $H=2.8$. A remarkable finding is that
local magnetizations of each sublattice are not uniform under the
competition of $E$ and $H$, i.e.,
$\vert \langle \sigma^x_{2i-1}\rangle\vert$ is slightly larger than
$\vert \langle \sigma^x_{2i}  \rangle\vert$, while
$\vert \langle \sigma^y_{2i-1}\rangle\vert$ and
$\vert \langle \sigma^y_{2i-1}\rangle\vert$ are smaller than their
counterparts. The ferrimagnetic structure is observed in some
magnetoelectric materials, such as hexaferrites
Ba$_{0.5}$Sr$_{1.5}$Zn$_2$Fe$_{12}$O$_{22}$ \cite{Momozawa93} and
Ba$_2$Mg$_2$Fe$_{12}$O$_{22}$ \cite{Ishiwata08}.

\begin{figure}[t!]
\includegraphics[width=8cm]{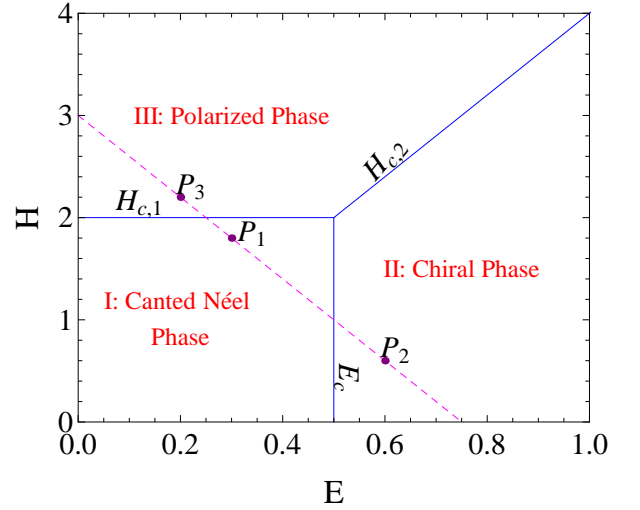}
\caption{(Color online)
Magnetic phase diagram of the 1D GCM at
$\theta=\pi/3$ with three critical lines are $H_{c,1}$, $H_{c,2}=4 E$
and $E_c$.
The magenta dashed line denotes a typical path $H=3-4 E$ that will be
used in the following discussion. Parameters and the representative
points $\{P_1,P_2,P_3\}$ in the $(E,H)$ plane are defined in
Fig. \ref{spectrum-k}.
As $\theta\rightarrow \pi/2$, the CN
phase collapses as both  $H_{c,1}$ and $E_c$ tend to zero.}
\label{PD-D-h}
\end{figure}

Following all the criteria including correlators, chirality, string
order parameter, entanglement spectrum and fidelity, the phase diagram
in the $(E,H)$ plane is displayed in Fig. \ref{PD-D-h}. For relatively
small $H$ and $E$ one finds region I which corresponds to the CN phase,
limited by two critical lines at $E=E_{c}$ and $H=H_{c,1}$.
The critical lines are defined by:
\begin{eqnarray}
H_{c,1}&=&2 \sqrt{J_o J_e} \vert \cos\theta \vert, \\
E_c&=& \frac{1}{2}\sqrt{J_o J_e} \vert \cos\theta \vert.
\end{eqnarray}
The area of the CN phase is proportional to $\cos^2 \theta$, and shrinks
to a point at $\theta=\pi/2$. While $E>E_c$, we find the third critical
line,
\begin{eqnarray}
H_{c,2}=4E,
\end{eqnarray}
which separates the chiral phase at low magnetic field from the
polarized phase at high magnetic field.

The long-range order is spoiled beyond the CN phase. Numerical results
show that $S_L$ saturates to a constant (\ref{SL}) in CN and polarized
phase and a logarithmic divergence (\ref{SL-L}) with $c=1/2$ is observed
along the critical line $H_{c,1}$, suggesting the QPT belongs to the 2D
Ising universality class. Specifically, $S_L$ also displays a
logarithmic form in the chiral phase, however, $c= 1$ is confirmed,
implying that the QPT to chiral phase falls within the well-known 1D XX
universality class. The logarithmic boundary-law violation is attributed
to its gapless nature. The gapless character will have a considerable
impact on the thermodynamic properties as any minor thermal fluctuation
should intermix the ground state and excited states.

\section{Thermodynamic properties}
\label{sec:Thermodynamic}

The remainder of the paper is concerned with the case where the system
is in thermal equilibrium. It is straightforward to obtain the
thermodynamic characteristics of the model Eq. (\ref{Hamiltonian1}) at
finite temperature. The free energy per site of the quantum-spin chain
at temperature $T$ is equal to
\begin{eqnarray}
{\cal F}=-\frac{T}{N} \sum_k \sum_{j=1}^{4}  \ln \left( 2 \cosh
\frac{\varepsilon_{k,j}}{T}\right).\label{thermodynamics}
\end{eqnarray}
Here we use the units with the Boltzmann constant set as $k_B\equiv 1$.
We derive the entropy which is arguably a fundamental thermodynamic
quantity and has been under consideration at low temperature since long
time ago \cite{Pauling35}. The entropy (${\cal S}$) and the specific
heat ($C_V$) --- we obtain both quantities from the free energy
${\cal F}$ via the standard relations:
\begin{eqnarray}
{\cal S}&=& -\frac{\partial{\cal F}}{\partial T}, \\
C_V     &=& -T\,\frac{\partial^2 {\cal F}}{\partial T^2}.
\end{eqnarray}

\begin{figure}[t!]
\includegraphics[width=8cm]{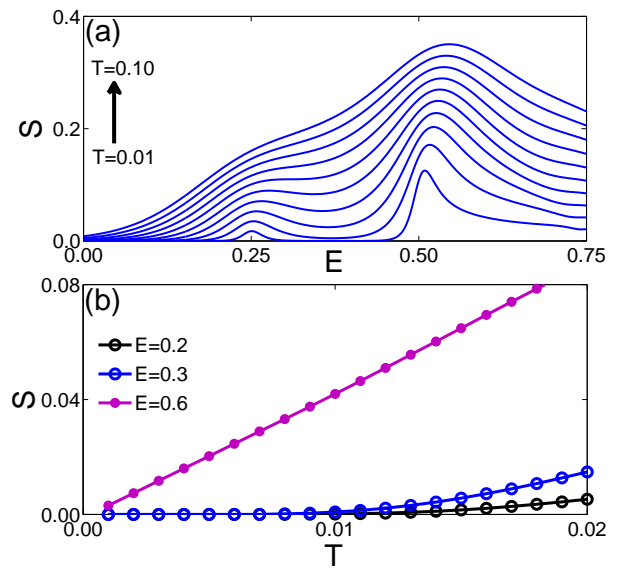}
\caption{(Color online)
Entropy ${\cal S}$ of the 1D $e_g$ orbital model at finite electric
field and temperature:
(a) the entropy versus electric field at increasing temperature
$T=0.01,0.02,\cdots,0.10$ (from bottom to top) along the path $H=3-4E$
defined in phase diagram shown in Fig. \ref{PD-D-h};
(b) scaling of entropy ${\cal S}(T)$ as a function of temperature $T$
for critical (filled circles, dashed line) and noncritical
(empty circles, solid line) fields.
Parameters are as follows: $J_o=1$, $J_e=4$, $\theta=\pi/3$.}
\label{S-T-h}
\end{figure}

As we show in Fig. \ref{S-T-h}(a), the entropy at low temperatures
displays two local maxima, implying two successive QPTs with the
increase of electric field $E$. One is close to $E=0.25$ and the other
is located at $E=0.5$. From the conformal field theory (CFT), the
low-temperature expansion of the free energy of the chiral phase per
site is given by \cite{Affleck86,Blote86,Cardy84},
\begin{eqnarray}
{\cal F}= \epsilon_0-\frac{\pi c}{6v_F}T^2 + {\cal O} (T^3),
\label{FTLL}
\end{eqnarray}
where $\epsilon_0$ is the ground state energy per site and $v_F$ is the
velocity of the excitations. Consequently, a linear relation of
${\cal S}$ with $T$ is observed in gapless chiral phase with scaling
\cite{Trippe10},
\begin{eqnarray}
{\cal S}=\frac{\pi c}{3v_F}T.
\end{eqnarray}
Here we adopt the units of $\hbar\equiv 1$.

The theoretical description of the chiral phase should be applicable
for the whole gapless regime in 1D systems. In this respect, the system
is gapless along the critical line and its entropy is also linear in
$T$, i.e., ${\cal S}(T)\propto T$, in the regime of low temperature,
while in the gapped phases an exponential scaling is observed, i.e.,
${\cal S}\propto\exp(-\Delta/T)$, see Fig. \ref{S-T-h}(b).

In fact, either in the chiral phase or at critical lines, the
low-temperature thermal entropy and the universal part of the
entanglement entropy are linked by a universal scaling function in the
framework of the 1D CFT, since both of them stem from the low-energy
degrees of freedom close to the Fermi surface. The dynamical critical
exponent $z$ controls the relative scaling of space and temperature
leading to an invariant form $L T^{1/z}$. For 1D relativistic
scale-invariant systems the finite-temperature fluctuations behave like
$(LT )^d$. Here we restrict $z=d=1$. On one hand, the von Neumann
entropy recovers the usual entanglement entropy of the ground state as
$L T^{1/z}\to 0$, and only constant or logarithmic terms are allowed at
$T=0$. A lot of studies revealed that the coefficient of the
boundary-law term (\ref{SL}) is nonuniversal, but the
boundary-law-violating term (\ref{SL-L}) at zero temperature was proven
to be universal. On the other hand, in the opposite limit as
$LT\to\infty$, the dimensionlessness and extensivity requires that the
thermal entropy per site then scales linearly with $T$ \cite{Swingle2013}.

\begin{figure}[t!]
\includegraphics[width=8cm]{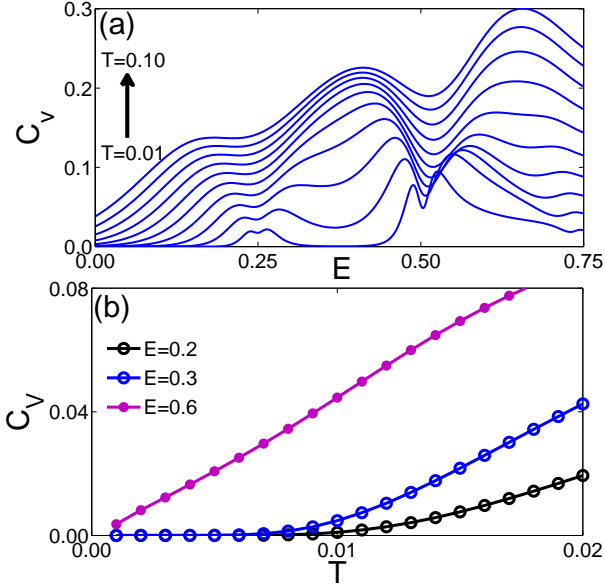}
\caption{(Color online)
Specific heat $C_V$ of the 1D $e_g$ orbital model as function of electric
field and temperature:
(a) the specific heat versus electric field along the path $H=3-4 E$
at different temperature $T=0.01,0.02,\cdots 0.10$ (from bottom to top)
the specific heat reaches its local minimums at QCPs for extremely low
temperatures;
(b) the scaling of specific heat at local minimum with respect to $T$.
Parameters are as follows: $J_o=1$, $J_e=4$, $\theta=\pi/3$.}
\label{Cv-T-h}
\end{figure}

The specific heat $C_V$ of the GCM in transverse field has been plotted
in Fig.\ref{Cv-T-h}(a). For extremely low temperatures, the specific
heat presents a broad peak around critical point and reaches a local
minimum on the top of peak at quantum critical points (QCPs). In the
gapped phase, the low-temperature specific heat reveals an exponential
increase in $T$ in the absence of spontaneous magnetization. Since
$C_V=T(\partial{\cal S}/\partial T)$, the specific heat at the QCPs
is identical to its entropy. In particular, the specific heat $C_V(T)$
of the Luttinger liquid is also linear in $T$ \cite{Giamarchi04}.

\begin{figure}[t!]
\includegraphics[width=8cm]{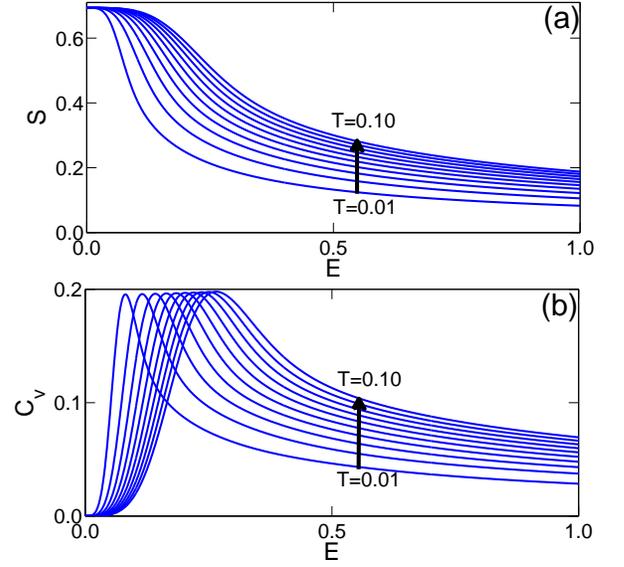}
\caption{(Color online)
Thermodynamic properties {\it of the QCM} for increasing electric field
$E$ as obtained along the {\it phase boundary} $H=4 E$:
(a) the entropy ${\cal S}$, and
(b) the specific heat $C_V$.
Different lines are for increasing temperature $T=0.01,0.02,\cdots,0.10$
(from bottom to top at $E=0.5$).
Parameters are as follows: $J_o=1$, $J_e=4$, $\theta=\pi/2$.}
\label{S_Cv-T-theta=pi.2}
\end{figure}

A special case is the quantum compass model (QCM) realized at $\theta=\pi/2$ .
Here we find that the low-temperature behavior is non-Fermi liquid like
and remarkably different from the behavior obtained at other values of
$\theta$. The entropy at different temperatures is plotted as function
of the $E$-field in Fig. \ref{S_Cv-T-theta=pi.2}(a) along the critical
line $H=4E$ which extends here to $E=0$. We recall, that for
$\theta=\pi/2$ the CN phase has disappeared. Usually, according to the
third law of thermodynamics, the entropy falls to zero at $T\to 0$.
Here it approaches the maximal value ${\cal S}=\ln 2$ per unit cell for
small $E$ field in the low-$T$ limit, as a result of the macroscopic
degeneracy $2^{N/2-1}$ in the disordered state \cite{You07}. Large
residual entropy was measured in the spin ice system Dy$_2$Ti$_2$O$_7$
\cite{Matsuhira02} where it is related to a macroscopic degeneracy of
the ground state resulting from frustration in the pyrochlore lattice.

The measurement of the full magnetic field and temperature dependence
of complete entropic landscape was performed for Sr$_3$Ru$_2$O$_7$ near
quantum criticality \cite{Rost09}. Lowering the entropy of ultracold
gases becomes nontrivial to realize more exotic quantum states
\cite{Ho09,Catani09,McKay11}, such as $d$-wave superconductivity.
Simultaneously, the specific heat remains zero for vanishing field,
as shown in Fig. \ref{S_Cv-T-theta=pi.2}(b).
This follows from the excitation gap which opens at this value of
$\theta$ between the degenerate ground state and excited states.

\section{Magnetoelectric effects}
\label{sec:Magnetoelectric}

\begin{figure}[t!]
\includegraphics[width=8cm]{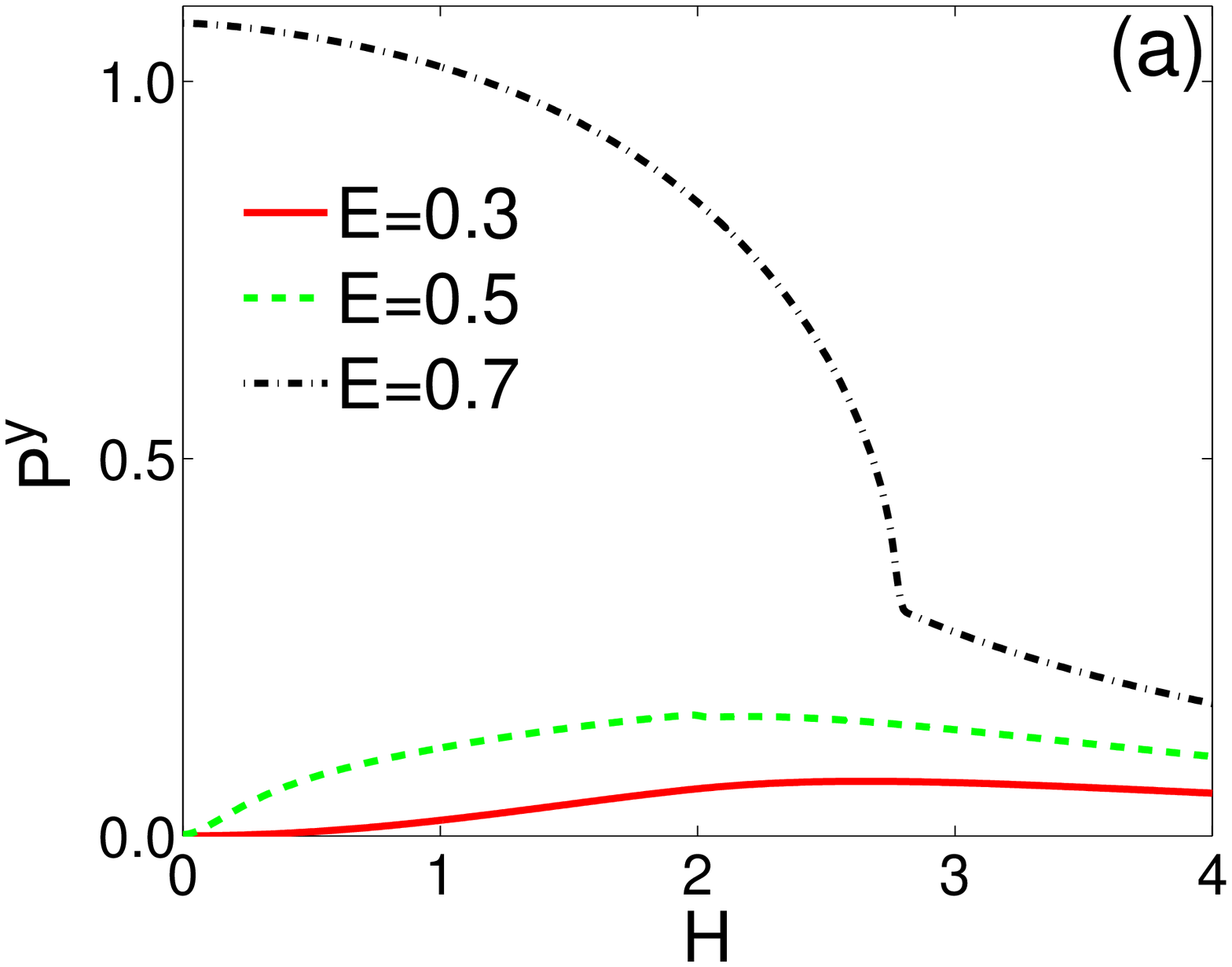}
\includegraphics[width=8cm]{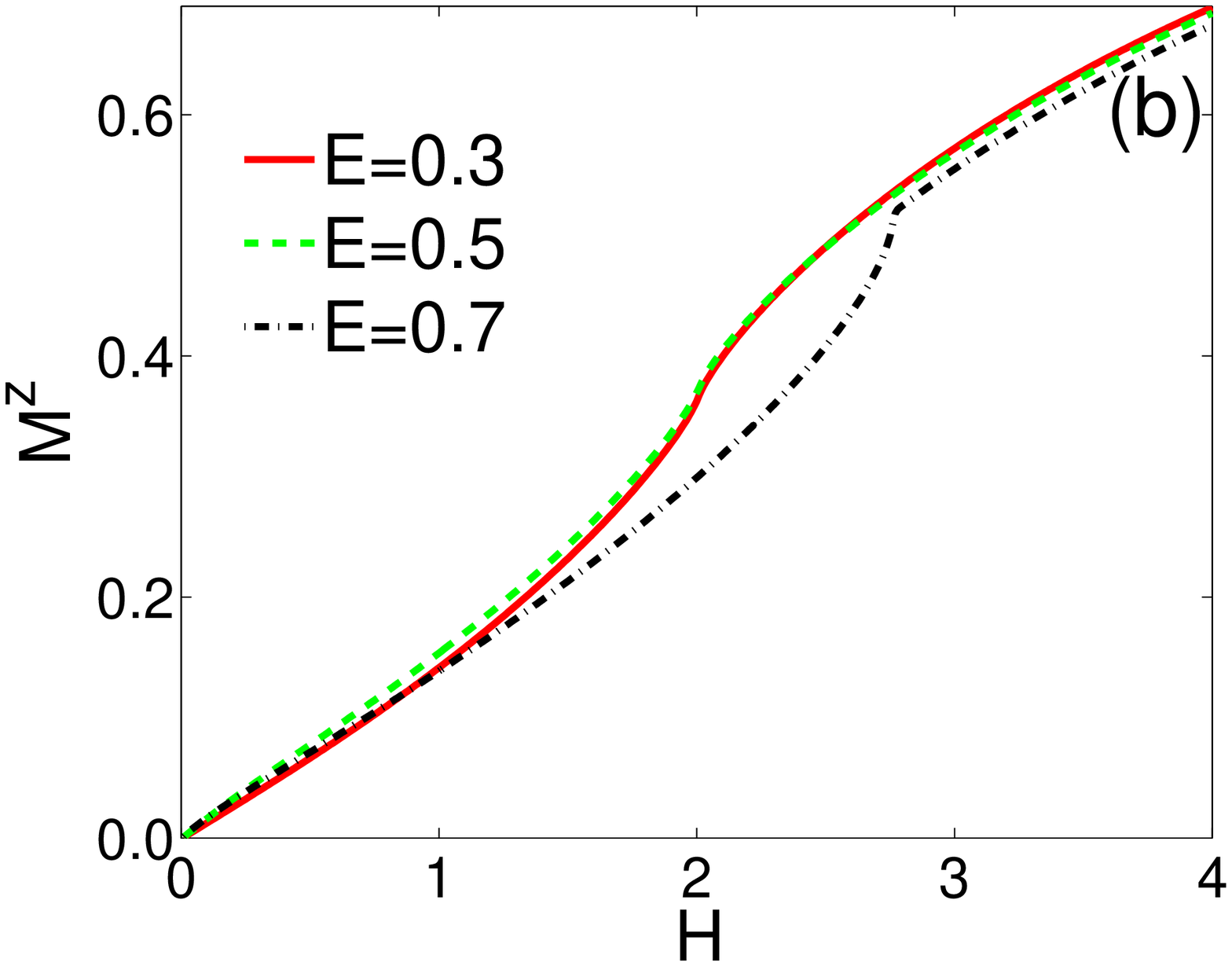}
\includegraphics[width=8cm]{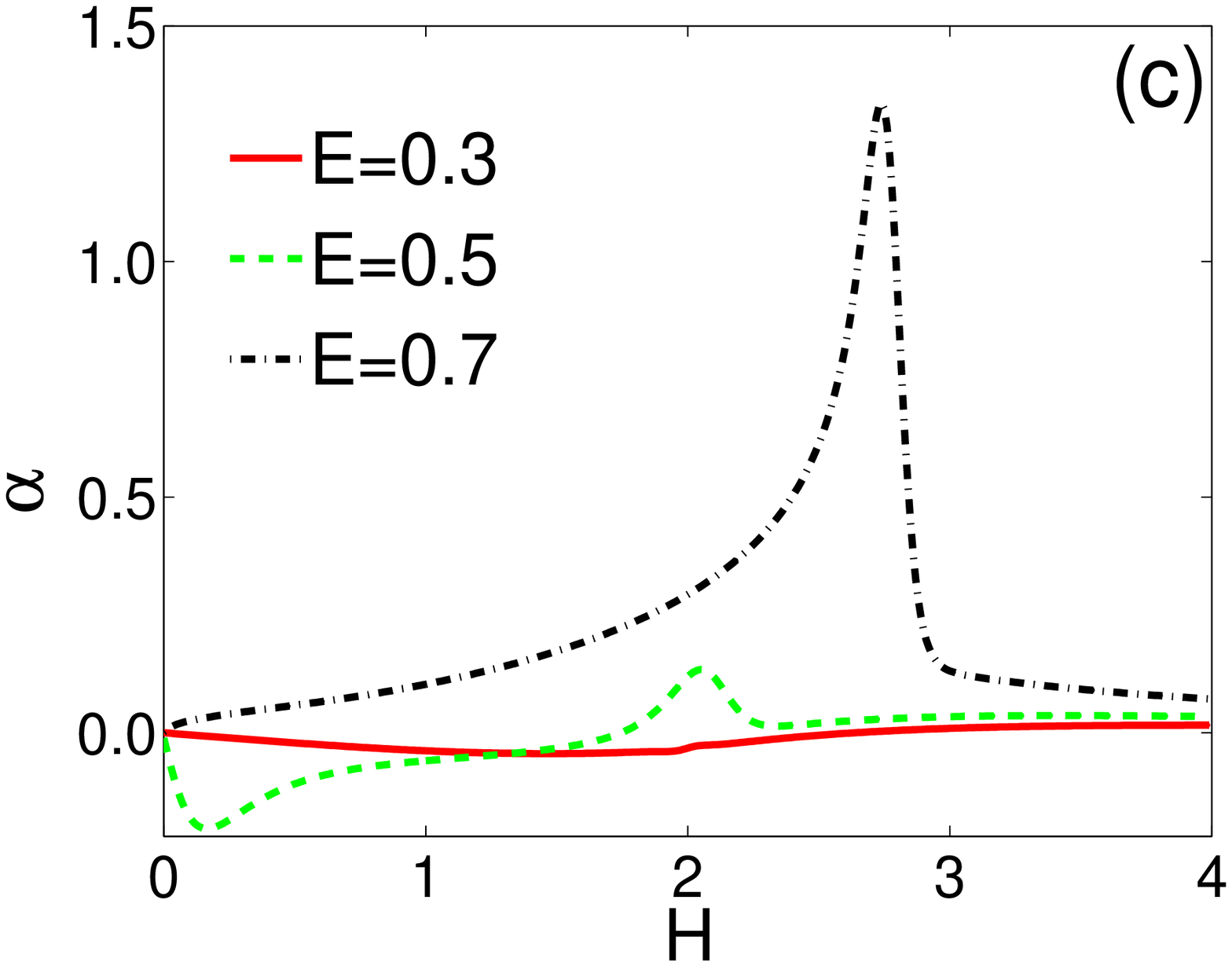}
\caption{(Color online)
Evolution of various quantities with increasing magnetic field $H$ at
low temperature $T=0.01$ and three values of electric field,
$E=0.3,0.5,0.7$ (see legend):
(a) $y$-direction electric polarization $P^y$,
(b) $z$-component magnetization $M^z$, and
(c) the magnetoelectric tensor $\alpha$ through relation
$\partial_{H}P^y$.
Parameters are as follows: $J_o=1$, $J_e=4$, and $\theta=\pi/3$.
}\label{combinedPMalpha-fixedJ0Je-D-h-2}
\end{figure}

The advantage of the presented formalism is that the magnetization, the
electric polarization, and thereby the magnetoelectric tensor can be
calculated exactly for the entire temperature range relevant for the
phase diagram. Figure \ref{combinedPMalpha-fixedJ0Je-D-h-2} shows the
average magnetization,
\begin{equation}
M^z=\frac{1}{N}\sum_l\langle \sigma_l^z \rangle,
\end{equation}
and according to  Eq. (\ref{polarization}) and the effective electric
field specified in Eq. (\ref{E_parameter}) the resulting {\it electric polarization}
component is,
\begin{equation}
P^y=\frac{1}{N}\sum_l\langle \sigma_l^x\sigma_{l+1}^y-\sigma_l^y\sigma_{l+1}^x\rangle,
\end{equation}
as function of the magnetic field $H$  for a few selected values of the
electric field $E$ at very low temperatures. Here the angle brackets
$\langle\dots\rangle$ denote the thermal average.

A key quantity to characterize the MEE is the linear magnetoelectric
susceptibility at constant temperature. The numerical derivatives of
$P^y$ and $M^z$ with respect to $H$ and $E$ define the magnetoelectric
tensor $\alpha^{yz}$,
\begin{eqnarray}
\alpha^{\mu\nu}=-\left( \frac{\partial P^\mu}{\partial H^\nu}
\right)_{T,\vec{E}}=-\left( \frac{\partial M^\nu}{\partial E^\mu}
\right)_{T,\vec{H}}.
\end{eqnarray}
The size of the macroscopic MEE depends on the microscopic mechanism.
We recall that as consequence of the relation between the $E$-field
and the external field component $E^y$ specified in  Eq. (\ref{E_parameter})
we have
\begin{equation}
\alpha^{yz}=-\left( \frac{\partial M^z}{\partial E^y}\right)_{T,\vec{H}}
=-\gamma\left( \frac{\partial M^z}{\partial E}\right)_{T,\vec{H}},
\end{equation}
which highlights the dependence on the magneto-electric coupling
parameter $\gamma$.
Below we use the abbreviation $\alpha$ for
the magnetoelectric tensor component $\alpha^{y z}$.

In Fig. \ref{combinedPMalpha-fixedJ0Je-D-h-2}(a) the electric
polarization $P^y$ is large in chiral phase (case of $E=0.7$) and it
decreases strongly towards the phase transition to the polarized phase
at $H=H_{c,2}$. In contrast, in the CN phase (at $E=0.3$) $P^y$ starts
from zero and increases gradually with $H$ but remains small compared
to chiral phase. Actually we find that the polarization $P^y$ behaves
quadratically at small $H\simeq 0$ in the CN phase, and shows a gradual
decrease after entering the polarized phase. In the limit of large $H$
the polarization decreases again to zero.

In the CN phase the $M^z$ component of the magnetization [Fig.
\ref{combinedPMalpha-fixedJ0Je-D-h-2}(b)] grows with $E$ and reaches
a maximum at the critical line $E_c$, which is contrary to the trend
found at the critical line $H_{c,2}$ (not shown). Figure
\ref{combinedPMalpha-fixedJ0Je-D-h-2}(b) reveals the {\it {\rm s}-shape
continuous} variation of the magnetization $M^z$ at the phase
transition between the CN and polarized phase, i.e., for $E=0.3$ and
also for $E=0.5$. In contrast, the transition at $E=0.7$ appears as a
second order phase transition, where the $M^z$ order is suppressed
below $H_{c,2}$ by the appearance of the chiral order and the associated
strong variation of $P^y$. This naturally leads to a huge signal in the
MEE tensor component $\alpha$ as seen in Fig.
\ref{combinedPMalpha-fixedJ0Je-D-h-2}(c) for $E=0.7$. On the other hand,
$\alpha$ remains small in the CN and polarized phase at $E=0.3$, but
develops strong features at $E=0.5$ both at small $H$ and in the
vicinity of $H_c$.

\begin{figure}[t!]
\includegraphics[width=8cm]{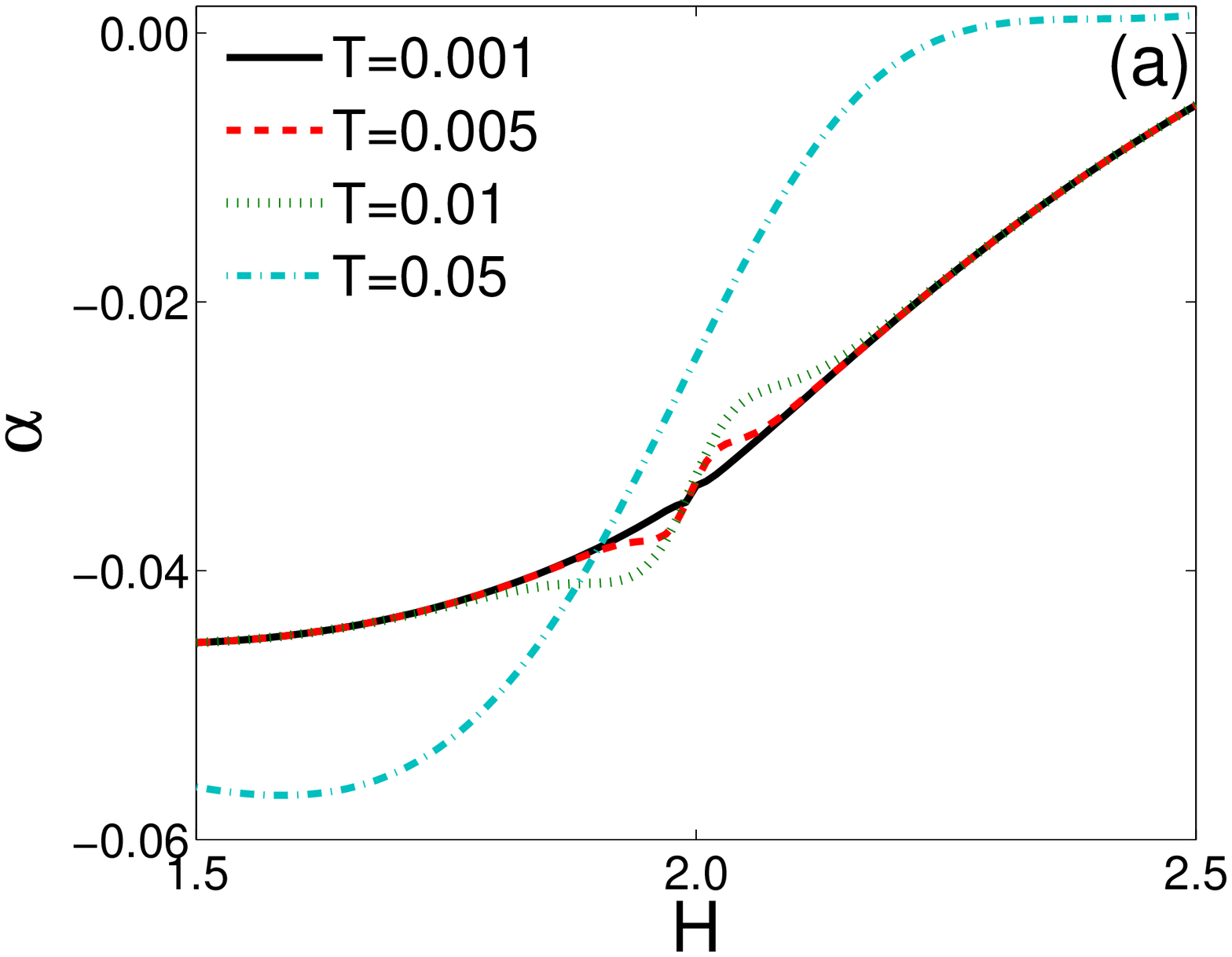}
\includegraphics[width=8cm]{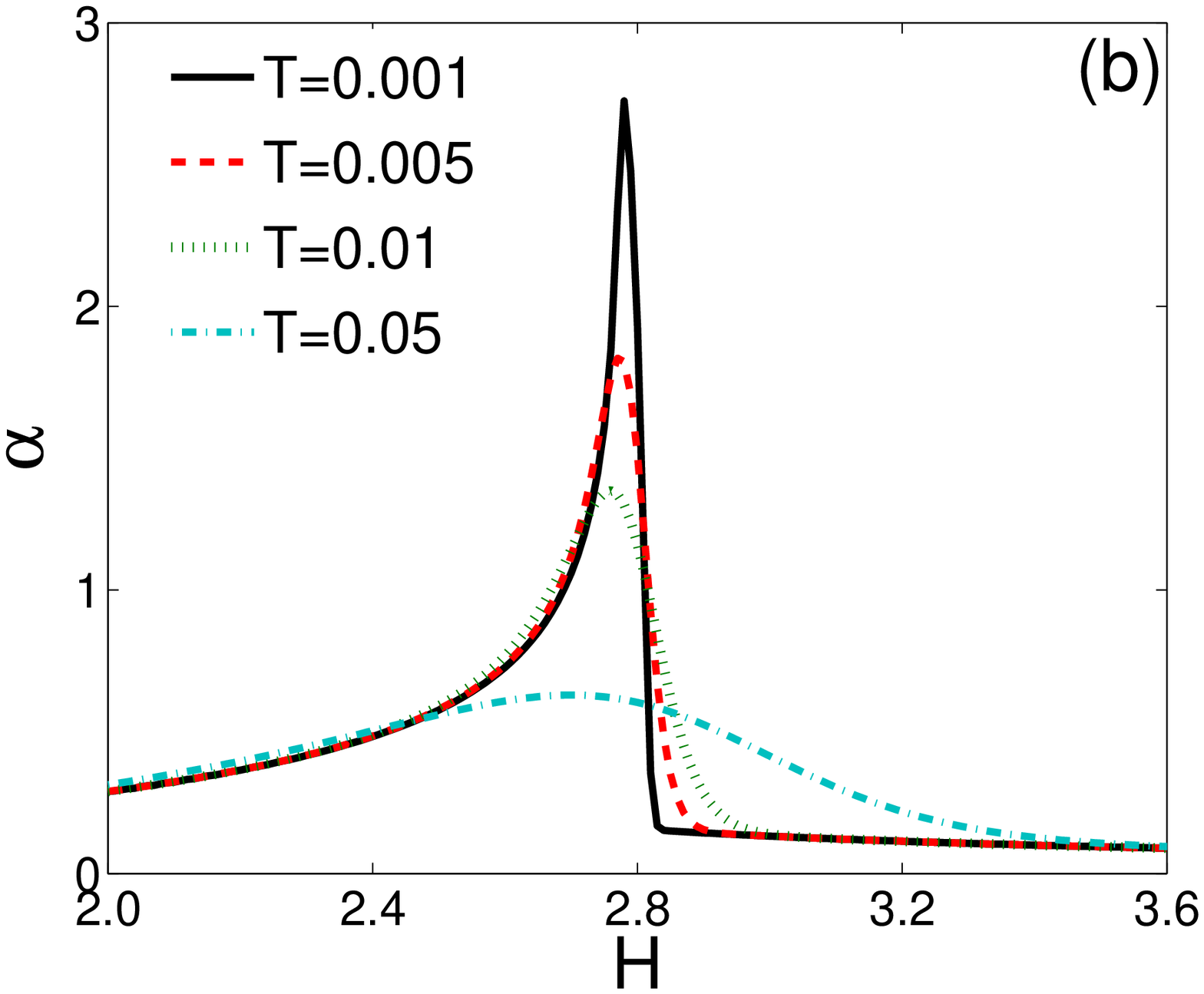}
\caption{(Color online)
Magnetoelectric tensor $\alpha$ as a function of $H$ obtained for:
(a) $E = 0.3$, and
(b) $E = 0.7$ close to the critical point.
The square-root singularity at $T=0$ decays rapidly with increasing
temperature.
Parameters are as follows: $J_o=1$, $J_e=4$, and $\theta=\pi/3$.}
\label{alpha-D-h-T}
\end{figure}

In  Fig. \ref{alpha-D-h-T} we compare the magnetic field dependence
of the magnetoelectric tensor $\alpha$ at different temperatures $T$.
It is evident that also the variation with temperature distinguishes:
(i) the phase transition from the CN to the polarized phase and
(ii)~the transition from the chiral to the polarized phase. In Fig.
\ref{alpha-D-h-T}(a) the magnetoelectric tensor undergoes a gradual
change at $E=2.0$ for $E=0.3$ under extremely low temperature. $\alpha$
manifests opposite trends on both sides of the critical point as the
temperature increases. Increasing temperature suppresses $\alpha$ in the
polarized state, while enhances it in the CN phase. Remarkably, $\alpha$
displays van Hove-like singularities close to $H_{c,2}=2.8$ for $E=0.7$.
These singularities gradually disappear at increasing temperature and
one finds that $\alpha$ becomes more and more flat, as shown in Fig.
\ref{alpha-D-h-T}(b). We have found that the singular behavior is
smeared out when $T>0.05$.

\begin{figure}[t!]
\includegraphics[width=8.2cm]{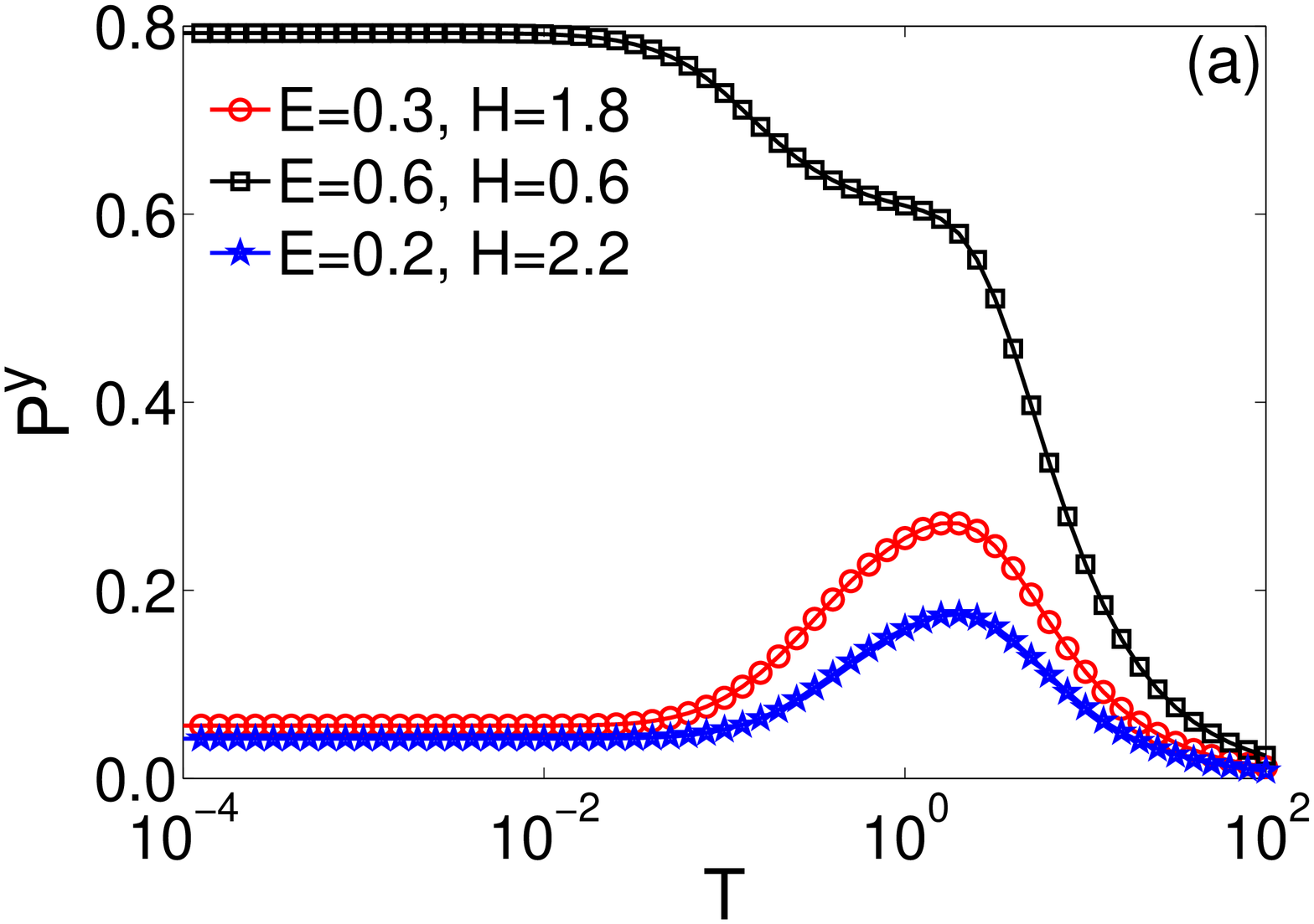}
\includegraphics[width=8.2cm]{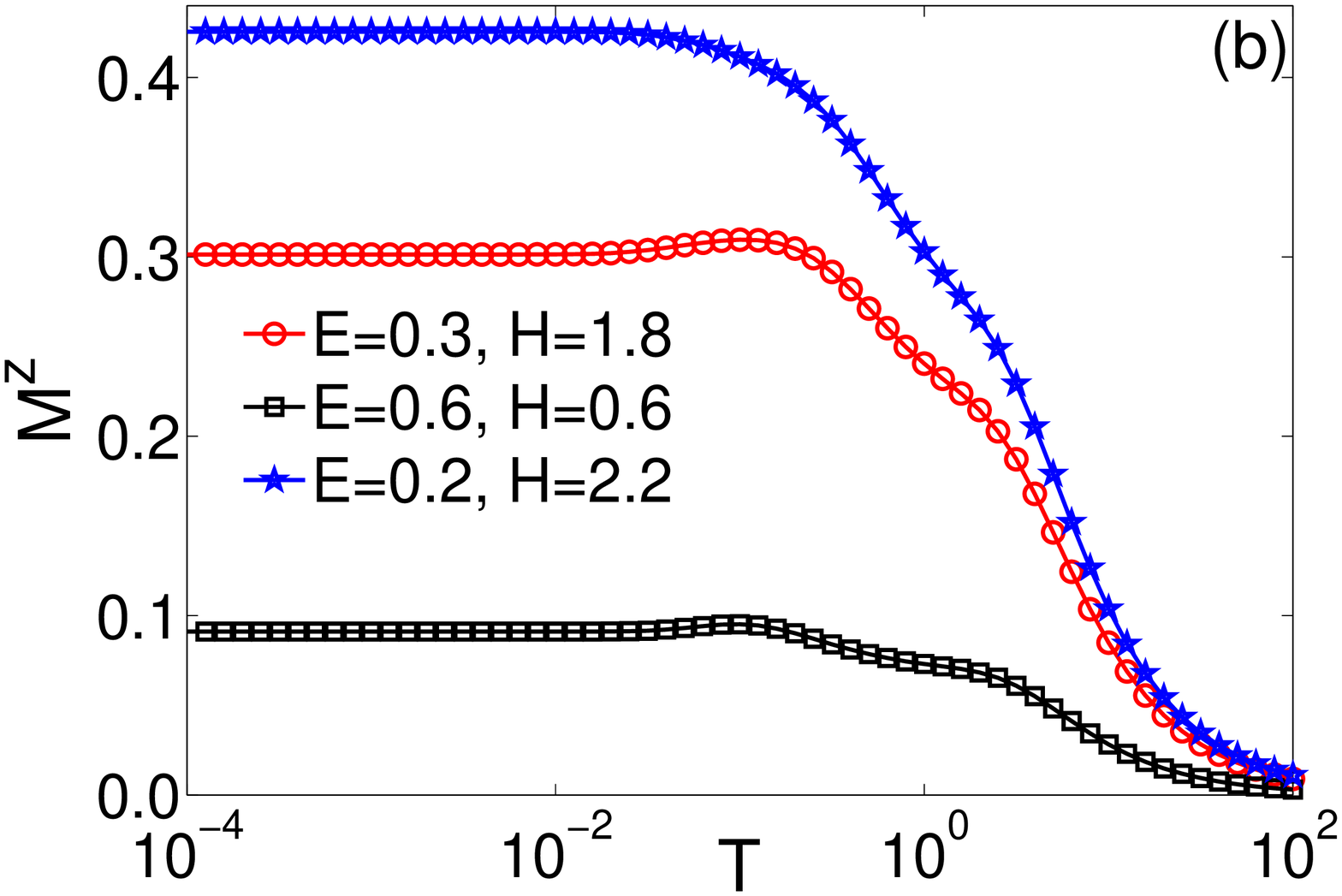}
\includegraphics[width=8.2cm]{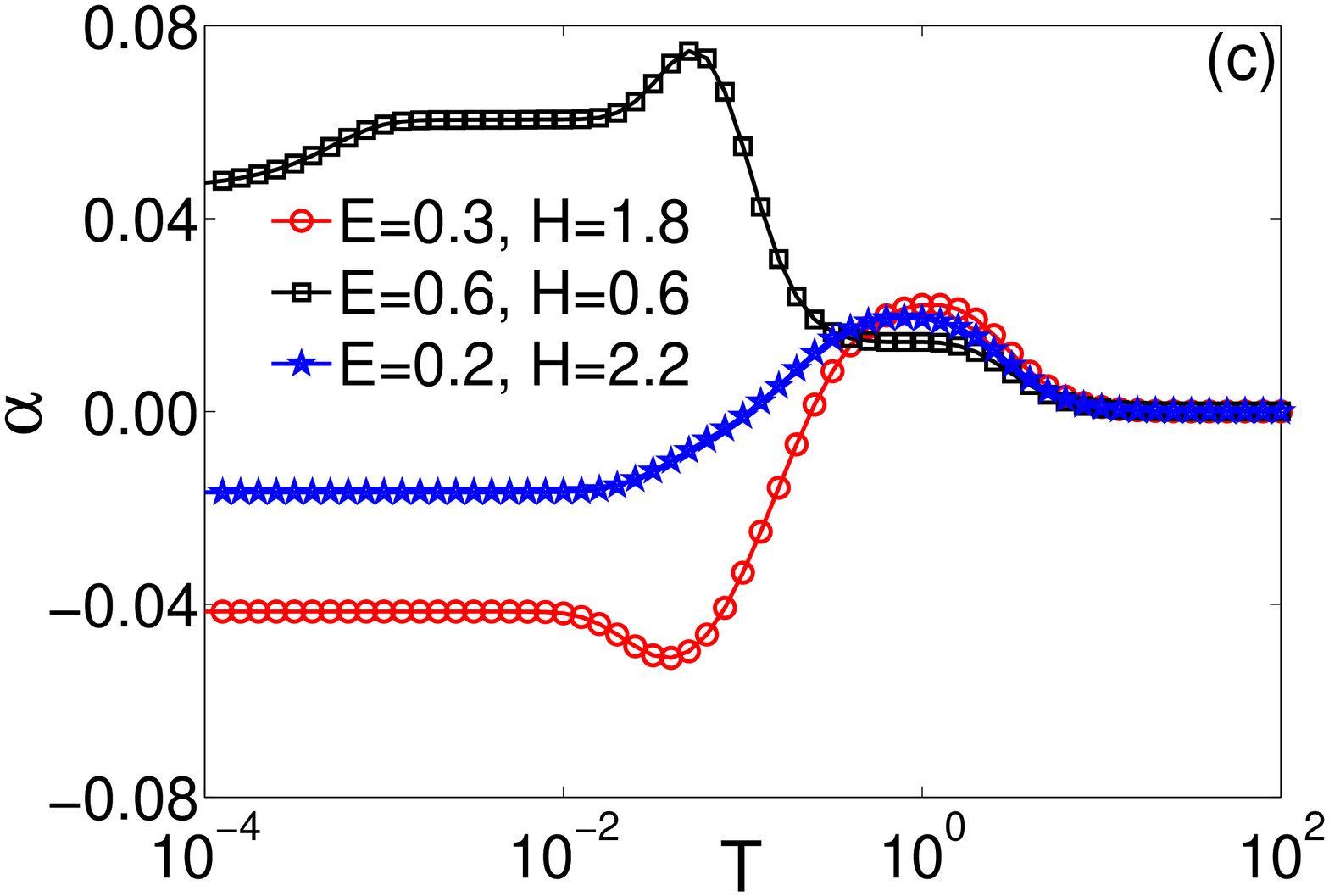}
\caption{(Color online)
Temperature dependence (note the logarithmic scale) of the
magnetoelectric effect for there selected $(E,H)$ points,
$P_1\equiv(0.3,1.8)$, $P_2\equiv(0.6,0.6)$, and $P_3\equiv(0.2,2.2)$,
corresponding to the CN, chiral and polarized phase, respectively:
(a) the polarization $P^y$,
(b) magnetization $M^z$,
(c) magnetoelectric tensor $\alpha$.
Parameters are as follows: $J_o=1$, $J_e=4$, and $\theta=\pi/3$.}
\label{alpha-T}
\end{figure}

The temperature dependence of $P^y$, $M^z$ and $\alpha$ is displayed
in Figs. \ref{alpha-T}(a)-\ref{alpha-T}(c). The data is shown at three
points $P_1$, $P_2$ and $P_3$, representing the CN phase, the chiral
phase and the polarized phase, respectively (see the phase diagram of
Fig. \ref{PD-D-h}). In Fig. \ref{alpha-T}(a) we find that the
$P^y$-component of the electric polarization for the point $P_2$
saturates at its maximal value at low temperatures.
With increasing temperature $P^y(T)$ decreases in two steps:
(i) the first decrease at $T_1 \sim 0.1$ can be identified with the
excitation energy between the chiral low energy states and nonchiral
excited states, while
(ii) the final decay of $P^y(T)$ towards zero at $T_2\sim 10$ can be
related to the total range of excitation energies.

Interestingly, the two characteristic temperature scales are also
recognized in the other phases. In the polarized and CN~phases $P^y(T)$
increases above $T_1$ from very small values at low temperatures and
assumes relatively large values near $T\sim 1$, i.e., comparable to
those in the chiral phase, and becomes small above $T_2$. Surprisingly,
the magnetoelectric effects are of similar strength in all three
magnetic phases in the intermediate temperature regime, $T_1<T<T_2$.

The same two characteristic temperatures may be recognized in the
temperature dependence of the magnetization $M^z(T)$ and
magnetoelectric tensor $\alpha(T)$.
The magnetoelectric tensor $\alpha$ changes sign from negative
to positive values upon increasing temperature reflecting the maximum
in $P^y(T)$ within the polarized and the CN phase. The derivative
$(\partial M^z/\partial T)$ also changes its sign at $T_{1}$. We note
that $T_{1}$ decreases monotonously when approaching chiral phase
along the path depicted in Fig. \ref{PD-D-h}, and the reentrant
behavior of $P^y$ vanishes after entering the chiral phase.

\section{DISCUSSION AND CONCLUSIONS}
\label{sec:discussion}

In this paper we considered the 1D generalized compass model which
interpolates between the Ising model ($\theta=0$) and the maximally
frustrated QCM ($\theta=\pi/2$) via the $e_g$ orbital
model, and includes Dzyaloshinskii-Moriya interaction. We investigated
this model in the presence of external magnetic and electric fields.
The Ising-like exchange interactions are directional in the compass
model and we selected the preferential axes in such a way that
interactions lie within the $(\sigma^x,\sigma^y)$ plane. The particular
advantage of the presented model is that it can be solved exactly in
terms of Jordan-Wigner transformation, and therefore the magnetization,
correlation functions, and the phase diagram could be obtained
rigorously. Note that usually the effective Ising interactions have to
be treated in the mean-field approximation which is uncontrolled.

The analytical results show that the angle $\theta$ between the easy
axes on odd and even bonds plays a crucial role in determining the
properties of the generalized compass model, including intersite spin
correlations and the excitation gap. We have shown that for
$\theta=\pi/3$, corresponding to the $e_g$ orbital model, the 1D model
is in the same universality class as the Ising model whose ground state
is N\'eel ordered. The presence of coplanar staggered
$\langle\sigma_i^{x(y)}\rangle$ order in this phase opens a possibility
for the existence of transverse order in addition. Indeed, the magnetic
field polarizes the system into ferromagnetic alignment. These two
phases exhaust the phase diagram at vanishing (or small) electric field.

In contrast, finite electric field drives the system into a chiral
phase, which is characterized by nonlocal $z$-th component string
order. The development of such peculiar phase can be regarded as a
spontaneous generation of Dzyaloshinskii-Moriya interaction, which
breaks the parity symmetry and exhibits a magnetoelectric effect. We
have demonstrated that the entanglement spectra of a half-infinite chain
as a function of electric field may also be used to determine the phase
diagram. Both N\'eel and polarized phases are gapped, where entanglement
is a constant satisfying the boundary law, while entanglement in the
gapless chiral state shows a logarithmic divergence.

By analyzing exact results at finite temperature obtained for entropy
and specific heat, we have established that the thermal properties
exhibit anomalies in the vicinity of quantum critical points. As a
function of the electric field, the entropy displays local maxima
while the specific heat exhibits local minima at critical points for
extremely low temperature, where a linear scaling with temperature was established. Away from the quantum critical point,
an exponential decay of the entropy and specific heat with the inverse temperature
is observed instead of a linear dependence on $T$ in the chiral phase.

The QCM is a very special case, which is realized at
the angle $\theta=\pi/2$, for which the spin components of the exchange
interactions along the even/odd bonds are orthogonal. The QCM
represents a peculiar quantum critical state between two gapped phases.
Removing external fields, i.e., at $E=0$ and $H=0$, the low energy
elementary excitations in Eq. (\ref{excitationspectrum}) become
dispersionless and tend to zero energy, i.e.,
$\varepsilon_{k,2}=\varepsilon_{k,3}=0$.
This flat band is then
half-filled by fermions, and thus gives high macroscopic degeneracy
$2^{N/2-1}$ away from the isotropic point, and the increased degeneracy
of $2^{N/2}$ when the spin interactions are balanced (at $J_o=J_e$).
The degeneracy for isotropic spin interactions increases further by a
factor of 2 in the thermodynamic limit, being $2\times 2^{N/2}$.

The critical lines intersect at $\theta_c=\pi/2$, $H_{c,1}=0$, $E_c=0$,
forming a multicritical point. The phase diagram changes qualitatively
in this case. The $z=2$ critical Fermi surface corresponds to a marginal
Fermi liquid, and it has a non-zero entropy density as $T\to 0$. This
indicates the absence of N\'eel-like long-range order in the ground
state of the $\theta=\pi/2$ compass model. A finite magnetic field
opens an exponentially small gap at the Fermi energy and thus removes
the high degeneracy of the ground state. However, applying the
Dzyaloshinskii-Moriya-type electric field, the spectra remain gapless at
$k=0$, but the huge degeneracy is lifted. The gap is then much smaller
than the external fields and therefore the thermal excitations through
the gap contribute to the thermodynamic properties at relatively low
temperature. This is observed in the maximal entropy ${\cal S}=\ln 2$
per unit cell being robust for not too large external field, as shown
in Fig. \ref{S_Cv-T-theta=pi.2}. The high degeneracy revealed by finite
entropy at low temperature suggests that the $\theta=\pi/2$ compass
model may have potential applications in quantum computation
\cite{Rio11}. In contrast, the entropy ${\cal S}$ of Fermi liquid
vanishes at zero temperature for $\theta\neq \pi/2$ according to the
third law of thermodynamics.

In summary, we have shown that the polarization as a function of
electric field is strongly affected by the magnetic field. Similarly,
the electric field has an effective impact upon the magnetization, which
depends on the strength of the magnetic field. Strong variation of
correlation functions and thermodynamic quantities are encountered by
varying both electric and magnetic field in the vicinity of a quantum
critical point, where the magnetoelectric tensor demonstrate
singularities at zero temperature. Remarkably, two characteristic
temperature scales are uncovered. For $T<T_1\approx 0.05$, the
magnetization saturates in all three phases at large values, similar to
the electric polarization $P^y$ in the chiral phase, whereas in the
other two phases $P^y$ drops to small (but finite) values. Within the
intermediate temperature range, $T_1<T<T_2$, the thermal excitation
admixes the features of chiral state and nonchiral state. As long as
the thermal energy overcomes the bandwidth for $T>T_2$, the high
temperature will wipe out the chiral features. This leads to a
characteristic reentrant behavior of the electric polarization $P^y$ in
the canted N\'{e}el and the ferro-polarized phases.

Moreover our work sheds light on the topological phase transitions
in strongly correlated systems through the mapping of the complex
spin Hamiltonian into the framework of  independent electrons.
Thereby the topological phase transition between the
canted N\'{e}el or the polarized  phase and the chiral phase, respectively,
acquires a different and perhaps simpler interpretation than in the original
spin model.

\acknowledgments

W.L.Y. acknowledges the helpful discussion with R. Tang.
This work is supported by the National Natural Science Foundation of China
(NSFC) under Grants No. 11474211 and No. 11347008 and the Natural Science
Foundation of Jiangsu Province of China under Grant No. BK20141190.
A.M.O. acknowledges support by the Polish National Science Center
(NCN) under Project No. 2012/04/A/ST3/00331.

\end{document}